\documentclass[pra,twocolumn,,amsmath,amssymb,floatfix]{revtex4}
\usepackage{graphicx}
\usepackage{bm}
\usepackage{amsmath,amssymb}

\def \be{\begin{equation}}
\def \ee{\end{equation}}
\def \ba{\begin{array}}
\def \ea{\end{array}}
\def \beq{\begin{eqnarray}}
\def \eeq{\end{eqnarray}}

\def \bed{\begin{displaymath}}
\def \eed{\end{displaymath}}
\def \no{\nonumber}

\newcommand{\HH}{{\cal H}}

\newcommand{\lb}{\left[}
\newcommand{\rb}{\right]}
\newcommand{\lp}{\left(}
\newcommand{\rp}{\right)}

\begin{document}
\title{Adiabatic perturbation theory: from Landau-Zener problem to quenching through a quantum critical point}

\author{C. De Grandi and A. Polkovnikov}
\affiliation{Department of Physics, Boston University, 590 Commonwealth Avenue, Boston, MA 02215, USA}

\begin{abstract}

We discuss the application of the adiabatic perturbation theory to analyze the dynamics
in various systems in the limit of slow parametric changes of the Hamiltonian. We
first consider a two-level system and give an elementary derivation of the asymptotics of the transition probability when the tuning parameter slowly changes in the finite range. Then we apply this perturbation theory to many-particle systems with low energy spectrum characterized by quasiparticle excitations. Within this approach we derive the scaling of various quantities such as the density of generated defects,  entropy and energy. We discuss the applications of this approach to a specific situation where the system crosses a quantum critical point. We also show the connection between adiabatic and sudden quenches near a quantum phase transitions and discuss the effects of quasiparticle statistics on slow and sudden quenches at finite temperatures.

\end{abstract}
\maketitle

\section{Introduction}

The dynamics in closed systems has recently attracted a lot of theoretical interest largely following the experimental developments in cold atoms systems~(see E.g. Ref.~\cite{Bloch2008_rmp} for a review). Several spectacular experiments already explored different aspects of the non-equilibrium dynamics in interacting many-particle systems~\cite{Greiner2002b, fertig2005, ari,  newtoncradle, sadler, schumm2007, Trotzky2008}. Recent theoretical works in this context have focused on various topics, for instance: the connection of dynamics and thermodynamics~\cite{olshanii_nature, reimann, silva, rigol2009_therm}, the dynamics following a sudden quench in low dimensional systems~\cite{ehud_assa, psg, barankov, cardy_quench1, cardy_quench2, powell, kollath, yuzbashyan, silva, roux, gritsev_quench,manmana, ludwig_quench, cazalilla_quench, barmettler}, the adiabatic dynamics near quantum critical points~\cite{ap_adiabatic, zurek_adiabatic, jacek, cherng-levitov, pg_np, ag, bp_optim, rafi, claudia, sengupta_sen2008a, sengupta_sen2008b, agkp, dutta2009, dutta2009a, sengupta_sen2009, itin, silva2009, moore2009}. Although there is so far very limited understanding of the generic aspects of the non-equilibrium quantum dynamics, it has been recognized that such issues as integrability, dimensionality, universality (near critical points) can be explored to understand the non-equilibrium behavior of many-particle systems in various specific situations.

The aim of this paper is to address some generic aspects of nearly adiabatic dynamics in many-particle systems. In particular, we will discuss in details the scaling of the density of generated quasiparticles $n_{\rm ex}$, entropy $S_d$, and heating $Q$ (non-adiabatic part of the energy), with the quenching rate. It has been already understood that this scaling is universal near quantum critical points~\cite{ap_adiabatic, zurek_adiabatic, cherng-levitov}, and more generally in low-dimensional gapless systems~\cite{pg_np}. The universality comes from the fact that if the system is initially prepared in the ground state or in a state with small temperature, then under slow perturbations very few transitions happen and the system effectively explores only the low energy part of the spectrum. This low energy part can be described by a small number of parameters characterizing some effective low energy theory (typically field theory). The situation can become different, however, in high dimensional systems~\cite{pg_np}. This is mainly due to a typically small density of low energy states, which for e.g. for free quasiparticles scales as $\rho(\epsilon)\propto \epsilon^{d/z-1}$, where $d$ is the dimensionality and $z$ is the dynamical exponent determining the scaling of energy $\epsilon$ with momentum $k$ at small $k$: $\epsilon(k)\sim k^z$. As a result the transitions to high energy states dominate the dynamics and the universality is lost. A similar situation happens for sudden quenches near quantum critical points~\cite{claudia_quench}. In low dimensions the excitations of low energy quasiparticles determine the (universal) scaling of various thermodynamic quantities. However, in high dimensions the transitions to the high energy states following the quench become predominant. In this case one can use  the ordinary perturbation theory or linear response which predicts that $n_{\rm ex}, S_d,$ and $Q$ become analytic (quadratic) functions of the rate for slow quenches and of the quench amplitude for sudden quenches.

 In this paper we will explain in details how the transition between quadratic and universal regimes can be understood as a result of breakdown of the linear response. More specifically we will illustrate how exactly the crossover between different scaling regimes occurs in the situations where the system can be well described by quasiparticle excitations. We will concentrate on the slow, linear in time, quenches and briefly mention the situation with fast quenches in the end. First we will discuss the adiabatic perturbation theory and its implications for many-particle systems. Then using this theory we will analyze a simple driven two-level system (Landau-Zener problem~\cite{lz1, lz2}) where the coupling linearly changes in time in the finite range. We will show how the quadratic scaling of the transition probability with the rate emerges from this perturbation theory. Then we will consider a more complicated situation where the system consists of free gapless quasiparticle excitations. We will show that in low dimensions, $d\leq 2z$, the scaling of the density of excitations and entropy is universal $n_{\rm ex}, S_d\sim |\delta|^{d/z}$, while in high dimensions the quadratic scaling is restored. The quadratic scaling can be understood as the result of multiple Landau-Zener transitions to high energy quasiparticle states. We will illustrate our argument with a specific model of coupled harmonic oscillators. Next we will consider a more complicated situation where the system is quenched through a second order quantum phase transition. We will show how the universal scaling $n_{\rm ex}\sim |\delta|^{d\nu/(z\nu+1)}$~\cite{ap_adiabatic, zurek_adiabatic} ($\nu$ is the critical exponent for the correlation length) emerges from combining adiabatic perturbation theory and universal scaling form of energies and matrix elements near the quantum critical point. We will also show how this scaling law breaks down and is substituted by a simple quadratic relation $n_{\rm ex}\sim \delta^2$ when the exponent $d\nu/(z\nu+1)$ exceeds two. We will illustrate these results using specific exactly solvable models. We will also discuss the connection between adiabatic and sudden quenches near the quantum critical point. In particular, we will show that in low dimensions, $d\nu<2$, the density of excited quasiparticles for a slow quench can be understood as a result of a sudden quench, if one correctly identifies the quench amplitude $\lambda^\star$ (for the sudden quench) with the quench rate $\delta$ (for the slow quench): $\lambda^\star\sim |\delta|^{1/(z\nu+1)}$. This analogy is very similar to the Kibble-Zurek picture of topological defect formation for quenches through classical phase transitions~\cite{kz1, kz2}, where one assumes that below certain energy (temperature) scale topological excitations essentially freeze. However at higher dimensions, $d\nu>2$, this analogy becomes misleading since for sudden quenches scaling of $n_{\rm ex}$ is no longer determined by low energy excitations. This work mostly focuses on the situation in which the system is initially in the ground state. In the end of the paper we will discuss what happens if the system is initially prepared at finite temperature. We will argue that the statistics of low energy quasiparticles strongly affects the response of the system to fast or slow quenches,  enhancing the non-adiabatic effects (compared to the zero-temperature case) in the bosonic case and suppressing them in the fermionic case. We will discuss the corrections to the universal scaling laws if the low energy quasiparticles are described by either bosonic or fermionic statistics.

\section{Adiabatic Perturbation Theory}
\label{Sec:pert}

We consider the following setup: the system is described by the Hamiltonian
$\mathcal H(t)=\mathcal H_0+\lambda(t)V$, where $\mathcal H_0$ is the
stationary part and $\lambda(t) V$ is the time-dependent part
of the Hamiltonian. Our purpose is to characterize the dynamics of this system
resulting from the time-dependent perturbation. We consider the
situation in which the system is in a pure state. More general situations, where
the state is mixed, can be addressed similarly by either solving von Neumann's
equation or averaging solutions of the Schr\"odinger equation with respect to
the initial density matrix. We assume that $\lambda(t)$ is a
linear function of time:
\be
\lambda(t)=\left\{\begin{array}{ll}
\lambda_i & t<0\\
\lambda_i+t \delta (\lambda_f-\lambda_i) & 0\leq t\leq 1/\delta\\
\lambda_f & t>1/\delta
\end{array}\right. .
\label{lin_quench}
\ee
Here $\delta$ is the rate of change of the parameter $\lambda(t)$: $\delta\to 0$
corresponds to the adiabatic limit, while $\delta\to\infty$ corresponds to a
sudden quench. In principle the values $\lambda_i$ and $\lambda_f$ can be arbitrarily far
from each other, therefore we can not rely on the conventional
perturbation theory in the difference between couplings $|\lambda_f-\lambda_i|$.

In the limit of slow parametric changes, we can use $\delta$
as a small parameter and find an approximate solution of the Schr\"odinger
equation:
\be
i\partial_t |\psi\rangle =\mathcal H(t) |\psi\rangle,
\label{schr_eq}
\ee
where $|\psi\rangle$ is the wave function. Here we  use the convention that $\hbar=1$ (this can be always achieved by rescaling either energy or time units). Our analysis will be similar to the one in Ref.~\cite{ortiz2008}, nevertheless for completeness we will present here the details of the derivation. It is convenient to rewrite the Schr\"odinger equation (\ref{schr_eq}) in the adiabatic (instantaneous) basis:
\be
|\psi(t)\rangle=\sum_n a_n(t) |\phi_n(t)\rangle,
\ee
where $|\phi_n(t)\rangle$ are instantaneous eigenstates of the Hamiltonian $\mathcal H(t)$:
\be
\mathcal H(t) |\phi_n(t)\rangle=E_n(t)|\phi_n(t)\rangle
\ee
corresponding to the instantaneous eigenvalues $E_n(t)$. These eigenstates
implicitly depend on time through the coupling $\lambda(t)$. Substituting this
expansion into the Schr\"odinger equation and multiplying it by $\langle
\phi_m|$ (to shorten the notations we drop the time label $t$ in $|\phi_n\rangle$)
we find:
\be
i\partial_t a_m(t)+i\sum_n a_n(t) \langle \phi_m|\partial_t|\phi_n\rangle=E_m(t) a_m(t).
\ee
We then perform a gauge transformation:
\be
a_n(t)=\alpha_n(t) \exp\left[-i\Theta_n(t)\right],
\ee
where
\be
\Theta_n(t)=\int_{t_i}^t E_n(\tau)d\tau.
\ee
The lower limit of integration in the expression for $\Theta_n(t)$ is arbitrary. We chose it to be equal to $t_i$ for convenience. In consequence the Schr\"odinger equation becomes
\be
\dot\alpha_n(t)=-\sum_m \alpha_m(t) \langle n|\partial_t|m\rangle\exp\left[i(\Theta_n(t)-\Theta_m(t))\right].
\ee
which can also be rewritten as an integral equation:
\be
\alpha_n(t)=-\int_{t_i}^t dt' \sum_m \alpha_m(t') \langle n|\partial_{t'}|m\rangle\mathrm e^{i(\Theta_n(t')-\Theta_m(t'))}.
\label{int_eq}
\ee
If the energy levels $E_n(\tau)$ and $E_m(\tau)$ are not degenerate, the
 matrix element $\langle n|\partial_t|m\rangle$ can be written
 as
 \be
 \langle n|\partial_t|m\rangle=-{\langle n| \partial_t \mathcal H|
 m\rangle\over E_n(t)-E_m(t) }=-\dot\lambda(t){\langle n| V| m\rangle\over
 E_n(t)-E_m(t) },
 \ee
 where we emphasize that the eigenstates $|n\rangle$ and eigenenergies
 $E_n(t)$ are instantaneous.
If $\lambda(t)$ is a monotonic function of time then in Eq.~(\ref{int_eq})
one can change variables from $t$ to $\lambda(t)$ and derive:
\be
\alpha_n(\lambda)=-\int_{\lambda_i}^\lambda d\lambda' \sum_m \alpha_m(\lambda') \langle n|\partial_{\lambda'}|m\rangle\mathrm e^{i(\Theta_n(\lambda')-\Theta_m(\lambda'))},
\label{int_eq1}
\ee
where
\be
\Theta_n(\lambda)=\int_{\lambda_i}^\lambda d\lambda' {E_n(\lambda')\over \dot\lambda'}.
\ee
Formally exact Eqs.~(\ref{int_eq}) and (\ref{int_eq1}) can not be solved in the
general case. However, they allow for a systematic expansion of the solution
in the small parameter $\dot\lambda$. Indeed, in the limit $\dot\lambda\to 0$
all the transition probabilities are suppressed because the phase factors are
strongly oscillating functions of $\lambda$. The only exception to this
statement occurs for degenerate energy levels~\cite{LL3}, which we do
not consider in this work. In the leading order in $\dot\lambda$ only the term
with $m=n$ should be retained in the sums in Eqs.~(\ref{int_eq}) and
(\ref{int_eq1}). This term results in the emergence of the Berry phase~\cite{shankar}:
\be
\Phi_n(t)=-i\int_{t_i}^t dt' \langle n
|\partial_{t'}|n\rangle=-i\int_{\lambda_i}^{\lambda(t)} d\lambda' \langle n
|\partial_{\lambda'}|n\rangle,  \ee
so that
\be a_n(t)\approx
a_n(0)\exp[-i\Phi_n(t)].
\ee
In many situations, when we deal with real Hamiltonians, the Berry phase is
identically equal to zero. In general, the Berry phase can be incorporated into our formalism by doing a unitary transformation $\alpha_n(t)\to \alpha_n(t)\exp[-i\Phi_n(t)]$ and changing $\Theta_n\to \Theta_n+\Phi_n$ in
Eqs.~(\ref{int_eq}) and (\ref{int_eq1}).

We now compute the first order correction to the wave function assuming for
simplicity that initially the system is in the pure state $n=0$, so
that $\alpha_0(0)=1$ and $\alpha_n(0)=0$ for $n\neq 0$.
In the leading order in $\dot\lambda$ we can keep only one term with
$m=0$ in the sums in Eqs.~(\ref{int_eq}) and~(\ref{int_eq1}) and derive
\be
\alpha_n(t)\approx-\int_{t_i}^t dt' \langle n|\partial_{t'}|0\rangle\mathrm e^{i(\Theta_n(t')-\Theta_0(t'))},
\label{int_eq2}
\ee
 or alternatively
 \be
 \alpha_n(\lambda)\approx-\int_{\lambda_i}^\lambda d\lambda' \langle
 n|\partial_{\lambda'}|0\rangle\mathrm
 e^{i(\Theta_n(\lambda')-\Theta_0(\lambda'))}.
 \label{int_eq3}
 \ee
The transition probability from the level $|\phi_0\rangle$ to the level
$|\phi_n\rangle$ as a result of the process is determined by
$|\alpha_n(\lambda_f)|^2$.

The expression~(\ref{int_eq3}) can be further simplified in the case where the
initial coupling $\lambda_i$ is large and negative and the final coupling
$\lambda_f$ is large and positive, employing the stationary-phase
approximation. The complex roots of the equation $E_n(\lambda^\star)-E_0(\lambda^\star)=0$
define the stationary point. Consequently the dominant contribution to the
transition probability is determined by the negative imaginary part of the
phase difference $\Theta_n-\Theta_0$ evaluated at these
roots~\cite{LL3}:
\be
|\alpha_n|^2\propto \exp[-2\Im
(\Theta_n(\lambda^\star)-\Theta_0(\lambda^\star))].
\label{exp_prob}
\ee
In particular, for linearly changing coupling $\lambda(t)= \delta t$, we obtain
\be
|\alpha_n|^2\propto \exp\left(-{2\over \delta}\Im
\int^{\lambda^\star}[E_n(\lambda')-E_0(\lambda')] d\lambda'\right)
\ee
and the transition probability exponentially vanishes as $\delta\to 0$.

However there are many cases where the coupling $\lambda_i$ or
$\lambda_f$ or both are finite. In this case Eq.~(\ref{exp_prob}) is no longer valid
and the asymptotic values of the integrals in Eqs.~(\ref{int_eq2}) and
(\ref{int_eq3}) are determined by the initial and final times of
evolution. Using the standard rules for evaluating the integrals of fast
oscillating functions we find:
\begin{widetext}
\beq
 &&\alpha_n(t_f)\approx \left[i {\langle\phi_n|\partial_t|\phi_0\rangle\over E_n(t)-E_0(t)}-{1\over E_n(t)-E_0(t)}{d\over dt}{\langle\phi_n|\partial_t|\phi_0\rangle\over E_n(t)-E_0(t)}+\dots\right]\mathrm e^{i(\Theta_n(t)-\Theta_0(t))}\Biggr|_{t_i}^{t_f}\nonumber\\
 &&=\left[i \dot\lambda{\langle\phi_n|\partial_\lambda|\phi_0\rangle\over E_n(\lambda)-E_0(\lambda)}-\ddot\lambda{\langle\phi_n|\partial_\lambda|\phi_0\rangle\over (E_n(\lambda)-E_0(\lambda))^2}-\dot\lambda^2{1\over E_n(\lambda)-E_0(\lambda)}{d\over d\lambda}{\langle\phi_n|\partial_\lambda|\phi_0\rangle\over E_n(\lambda)-E_0(\lambda)}+\dots\right]\mathrm e^{i(\Theta_n(\lambda)-\Theta_0(\lambda))}\Biggr|_{\lambda_i}^{\lambda_f}.\phantom{XX}
 \label{expansion1}
 \eeq
\end{widetext}
In the following analysis we will retain only the first non-vanishing term in $\dot\lambda=\delta$.
The terms proportional to higher powers of the expansion parameter, such as $\ddot\lambda$, $(\dot\lambda)^2$, as well as non-analytic terms similar to Eq.~(\ref{exp_prob}), will be neglected assuming sufficiently small $\delta\to 0$.

The probability of the transition to the $n$-th level is approximated by:
\begin{widetext}
\be
|\alpha_n(\lambda_f)|^2\approx \delta^2\left[
{|\langle\phi_n|\partial_{\lambda_i}|\phi_0\rangle|^2\over (E_n(\lambda_i)-E_0(\lambda_i))^2}+{|\langle\phi_n|\partial_{\lambda_f}|\phi_0\rangle|^2\over (E_n(\lambda_f)-E_0(\lambda_f))^2}\right]-2\delta^2 {\langle\phi_n|\partial_{\lambda_i}|\phi_0\rangle\over E_n(\lambda_i)-E_0(\lambda_i)} {\langle\phi_n|\partial_{\lambda_f}|\phi_0\rangle\over E_n(\lambda_f)-E_0(\lambda_f)}\cos\left[\Delta\Theta_{n0}\right],
\label{tr_prob}
\ee
\end{widetext}
where
$\Delta\Theta_{n0}=\Theta_n(\lambda_f)-\Theta_0(\lambda_f)-\Theta_n(\lambda_i)+\Theta_0(\lambda_i)$
is the phase difference between the states $|\phi_n\rangle$ and
$|\phi_0\rangle$ accumulated during the time evolution. This phase difference
is usually very large and thus the last term in Eq.~(\ref{tr_prob}) is a
highly oscillating function, which can be typically dropped because of the
statistical or time averaging.

\subsection{Application to the Landau-Zener problem.}
\label{sec:lz}
We apply the general formalism presented above to the notorious Landau-Zener (LZ)
problem~\cite{lz1, lz2}. The Hamiltonian to study this problem is given by a $2\times 2$ matrix:
\be
\mathcal H=\lambda\sigma_z+g\sigma_x,
\ee
which is conveniently expressed through the Pauli matrices:
\be
\sigma_z=\left[\begin{matrix} 1 & 0 \\ 0 & -1\end{matrix}\right];\;
\sigma_x=\left[\begin{matrix}0 & 1 \\ 1 & 0\end{matrix}\right].
\ee
This system has two eigenstates:
\be
|-\rangle=\left(\begin{array}{c} \sin(\theta/2) \\ -\cos(\theta/2)\end{array}
\right),\quad |+\rangle=\left(\begin{array}{c} \cos(\theta/2) \\
\sin(\theta/2)\end{array}\right),
\label{eigen_lz}
\ee
where $\tan\theta=g/\lambda$, with corresponding  energies $E_{\pm}=\pm\sqrt{\lambda^2+g^2}$.

We assume that the coupling $\lambda$ linearly changes in time:
$\lambda=\delta t$. The system is initially, at $t=t_i$, prepared in the
ground state and the process continues until $t=t_f$. In the limit $t_i\to
-\infty$ and $t_f\to\infty$ the probability to occupy the excited state
$|+\rangle$~\cite{lz1, lz2} is a non-analytic function of $\delta$:
\be
|a_+|^2=\exp\left[-{\pi g^2\over \delta}\right].
\label{lz_exact}
\ee
However, in the general case, where $t_i$ or $t_f$ are finite, the probability
contains both non-analytic and analytic contributions in $\delta$.

In principle the LZ problem can be solved exactly for arbitrary $t_i$ and $t_f$~\cite{vitanov_96, vitanov_99} (Appendix~\ref{LZ_app}), however the general solution is quite cumbersome. Here we illustrate how the asymptotical behavior of the transition probability at small $\delta$ can be recovered employing the adiabatic perturbation theory.
The only non-zero matrix element, which enters Eq.~(\ref{int_eq3}), is
\be
\langle +|\partial_t|-\rangle=\dot\theta/2=-{1\over 2}{\dot\lambda g\over \lambda^2+g^2}.
\ee
 We first apply Eq.~(\ref{int_eq2}) to the case
$t_i\to-\infty$ and $t_f\to\infty$ which corresponds to the classic LZ
problem. Then Eq.~(\ref{int_eq3}) gives:
\be
\alpha_+(\infty)\approx {1\over 2}\int\limits_{-\infty}^\infty \, dt{\delta g\over g^2+(\delta t)^2}\exp\left[
2i\int_0^t d\tau\sqrt{(\delta \tau)^2+g^2}\right].
\label{eq:LZ_pert}
\ee
\begin{figure}[t]
\includegraphics[width=3.5in]{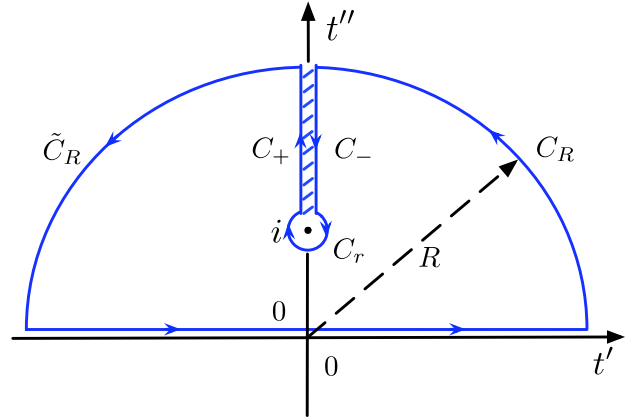}
\vspace{-0.5cm}
\caption[]{
The contour of integration in Eq.~(\ref{eq:LZ_pert}) in the complex $t$-plane.
Integration over the real axis is given by several contributions: the one
along $C_r$ around the point $t=i$ that gives the first term in Eq.(\ref{eq:LZ_int}), $C_+$ and $C_-$
from the two sides of the branch represented by second term in
Eq.(\ref{eq:LZ_int}), and also $C_R$ and $\tilde C_R$ that vanish in the limit
$R\to \infty$.}
\vspace{-0.5cm}
\label{fig:Higgs_diagram}
\end{figure}
The asymptotic behavior of this expression at small $\delta\ll g^2$ can be
derived by studying the analytic properties of the integrand in the complex
plane of the variable $t=t'+it''$. We notice that the phase factor has a
branch-cut singularity in the upper and lower-half planes along the imaginary
axis, that starts at $t''=\pm 1$ and goes to infinity at $t''=\pm\infty$.
Deforming the contour of integration to include a singularity, e.g. in the
upper-half plane, we find that the integral has two contributions: one is
provided by $t''=1$ point, and another one is given by a combination of two
paths that run from the complex infinity to $t''=1$ and backwards with the
corresponding phase shift (Fig. \ref{fig:Higgs_diagram}). As a result, we obtain:
\beq\label{eq:LZ_int}
&&\alpha_+(\infty)\approx \frac{\pi}{2}\exp\lb-\frac{\pi g^2}{2\delta}\rb  \\
&&\times \lp 1- \frac{2}{\pi}\Im\int_0^\infty\,\frac{dx}{\sinh x}\exp[\left[i \frac{g^2}{\delta}\left( x-\frac{1}{2}\sinh(2x)\right)\right]\rp,\nonumber
\eeq
where we changed the variables $t\delta/g=i\cosh x$ to simplify the
expressions. In the limit $\delta\ll g^2$, the integral in the brackets is a
constant equal to $\pi/6$, which leads to the following asymptotic behavior
of the transition probability:
\be
|\alpha_{+}(\infty)|^2\approx {\pi^2\over 9}\exp\left[-{\pi g^2\over
\delta}\right].
\label{lz_pert}
\ee
This expression correctly reproduces the exponential dependence of the
transition probability on the Landau-Zener parameter $g^2/\delta$. However,
the exponential prefactor is larger than the exact value of
unity~\cite{vitanov_96}. The reason for this discrepancy is that the exponential dependence is a result of a delicate interference of the transition amplitudes in time, which can not be obtained within perturbative approach. Conversely, as we argue below, in the case of finite $t_i$ or $t_f$, the asymptotical behavior of the transition probability with $\delta$ is analytic, allowing for a systematic treatment within the adiabatic perturbation theory.

Let us now turn to such situation with finite $t_i$ or $t_f$ or both (positive or negative) and sufficiently small $\delta$ . Then using Eq.~(\ref{tr_prob}) and ignoring the fast oscillating term we find
 \be
 \alpha_+(t_f)|^2\approx {\delta^2\over 16 g^4}\left({g^6\over (g^2+\lambda_i^2)^3}+{g^6\over (g^2+\lambda_f^2)^3}\right).
\label{tr_prob_lz}
\ee
In the case of $\lambda_i=-\infty$ and $\lambda_f=0$, i.e. $t_i=-\infty$ and $t_f=0$, this gives
 \be
 |\alpha_+(t_f=0)|^2\approx {\delta^2\over 16 g^4}
 \label{tr_prob_lz_sym}
 \ee
as it can be found solving exactly the LZ-problem (see Appendix \ref{LZ_app}).
We would like to emphasize that in agreement with the general
prediction~(\ref{tr_prob}) in the adiabatic limit the transition probability
(\ref{tr_prob_lz}) is  quadratic in the rate $\delta$.  We note that a more accurate asymptotic includes an additional exponential non-perturbative term~(\ref{lz_exact}) if $\lambda_i$ and $\lambda_f$ have
  opposite signs~\cite{vitanov_99}. With this additional term
 Eq.~(\ref{tr_prob_lz}) will give the correct asymptotic for $|\alpha_+(t_f)|^2$,
 even when both $|\lambda_i|$ and $|\lambda_f|$ are large.

\section{Adiabatic dynamics in gapless systems with quasiparticle
excitations}\label{gapless}

In the previous section we applied the adiabatic perturbation theory to a two-level system with a time-dependent gap separating the eigenstates of the Hamiltonian. Now we are interested in extending the analysis to the case of a many-particles system. If we consider a situation in which the system (initially prepared in the ground state) is characterized by gapped quasiparticle excitations, then clearly the previous analysis applies to each mode independently. Then under slow quench different quasiparticle states can be excited. Since the transition probability to each quasiparticle state quadratically depends on the quench rate $\delta$, we can expect that the density of created quasiparticles and other thermodynamic quantities will also quadratically depend on $\delta$. The situation becomes more complicated and more interesting if we consider a system with gapless excitations. Then for the low energy states the adiabatic conditions are effectively always violated (unless we consider quench rates which go to zero with the system size) and in principle the adiabatic perturbation theory and the quadratic scaling can break down. Although the adiabatic perturbation theory in this case is no longer quantitatively correct, here we will show that it can nevertheless  be very useful in finding the scaling of various quantities with the quench rate and  when this scaling becomes quadratic. Let us assume that we are dealing with a homogeneous system of quasiparticles characterized by a dispersion relation:
\be
\varepsilon_k=c(\lambda) k^z,
\label{disp}
\ee
where $z$ is the dynamical exponent and $c(\lambda)$ is a pre-factor depending on the external parameter $\lambda$. We assume that
$\lambda$ changes linearly in time, $\lambda=\delta t$,  between the initial and the final value
$\lambda_i$ and $\lambda_f$ respectively. The parameterization~(\ref{disp})
explicitly demonstrates that during the evolution the details of the
quasiparticle spectrum do not change, i.e. the system remains gapless and the
exponent $z$ stays the same. Another interesting possibility where the system
crosses a singularity such as a critical point, which violates this assumption,
will be considered in the next section. In this work we also limit ourselves to global uniform quenches, where the coupling $\lambda$ is spatially
independent.

In this section we consider the situation where the ground state corresponds to the state with no quasiparticles. Such situation naturally appears in a variety of physical systems, e.g.  bosonic systems with short-range
interactions, fermionic systems without Fermi surfaces such as gapless semiconductors~\cite{gelmont},  graphene~\cite{antonio_rmp} at zero voltage bias, some one-dimensional spin chains, that can be mapped to  systems of fermions using the Jordan-Wigner transformation~\cite{giamarchi_book, subir}, system of hard-core bosons in one-dimension (or Tonks gas)~\cite{subir, girardeaus, bloch_tonks} and so on. The situation with Fermi systems with the ground state corresponding to the filled Fermi sea requires special attention and will not be considered here. We do not expect any qualitative differences in the response to slow quenches in these Fermi systems.

Dimensional analysis allows us to estimate the scaling of various thermodynamic
quantities with the quench rate. On the one hand, in a gapless system the transitions to the low energy states are unavoidable since any change in the coupling $\lambda$ looks fast (diabatic) with respect to them. On the other hand transitions to the high energy states are suppressed for small $\delta$ because of the fast-oscillating phase factor entering the expression for the transition amplitude~(\ref{int_eq3}), so that the dynamics with respect to these states is adiabatic. It is straightforward to estimate the boundary separating diabatic and adiabatic states. Namely, if throughout the evolution $\dot\varepsilon_k(\lambda)$ becomes larger or comparable to $\varepsilon_k^2(\lambda)$, then the corresponding energy level is diabatic and the
quasiparticles are easily created. On the other hand if $\dot\varepsilon_k(\lambda)\ll \varepsilon_k^2(\lambda)$ in the whole interval of $\lambda\in [\lambda_i,\lambda_f]$ then the transitions are suppressed. Of course the implicit assumption here is that there is no kinematic constraint preventing creation of quasiparticles, i.e. that the matrix element for the transition is non-zero. In spatially uniform systems single quasiparticles typically can not be created because of the momentum conservation, so that quasiparticles can be created only in pairs with opposite momenta. This implies that one should use $2\varepsilon_k(\lambda)$ instead of $\varepsilon_k(\lambda)$ in the argument, however, this extra factor of two is not important for a qualitative discussion. The crossover energy $\tilde\varepsilon_k(\lambda)$ is found by equating the two expressions:
\be
\tilde{k}^z {\partial c(\lambda)\over \partial \lambda}\delta\sim c(\lambda)^2
\tilde k^{2z}
\ee
or alternatively
\be
\tilde\varepsilon_k(\lambda) \sim\delta {\partial \ln
c(\lambda)\over\partial\lambda};\quad \tilde k^z\sim \delta {1\over
c(\lambda)}{\partial \ln c(\lambda)\over\partial\lambda}.
\ee
More precisely one needs to find the maximal momentum satisfying the equation above in the whole interval of $\lambda\in[\lambda_i,\lambda_f]$. If there are no singularities during the evolution the derivative $\partial_\lambda \ln c(\lambda)$ remains non-zero and bounded. Thus we see that the crossover energy $\tilde \varepsilon_k(\lambda)$ scales linearly with the rate $\delta$. Then we can estimate, for example, the total number of quasiparticles created during the process as
\be
\label{d/z}
n_{\rm ex}(\delta)\sim\int_0^{\tilde \varepsilon}
\rho(\epsilon)d\epsilon=\int_{k\leq\tilde k} {d^d k\over (2\pi)^d}\propto
|\delta|^{d/z}.
\ee
We expect this scaling to be valid only when $d/z\leq 2$.  Otherwise (as we will argue below) the contribution from high energy quasiparticles with
$\varepsilon>\tilde\varepsilon$ will dominate, resulting in the quadratic scaling of $n_{\rm ex}$.

The result (\ref{d/z}) can be derived more accurately using the adiabatic perturbation theory. Namely, let us perform the scaling analysis of Eq.~(\ref{int_eq3}). Because quasiparticles can be typically created only in pairs with
opposite momenta one should use twice the quasiparticle energy in the dynamical phase $\Theta_n$ in Eq.~(\ref{int_eq3}). The number of quasiparticles should be also multiplied by two, but at the same time the sum over momenta should only go over half the available states to avoid double counting of pairs with momenta $k$ and $-k$. So this second factor of two can be absorbed extending the integration over momenta to the whole spectrum. Thus we get
\beq
&&n_{\rm ex}\approx {1\over L^d}\sum_k |\alpha_k(\lambda_f)|^2=\int {d^dk\over
(2\pi)^d}\nonumber\\
&&\left|\int_{\lambda_i}^{\lambda_f} d\lambda\,\langle
k|\partial_\lambda|0\rangle
\exp\left[{2ik^z\over\delta}\int_{\lambda_i}^\lambda c(\lambda')
d\lambda'\right]\right|^2.
\eeq
Rescaling the momentum $k$ as $k= \delta^{1/z}\eta$ we find
\be
n_{\rm ex}\approx |\delta|^{d\over z}\!\int {d^d\eta\over
(2\pi)^d}\left|\int\limits_{\lambda_i}^{\lambda_f} d\lambda\,\langle \eta
\delta^{1/z}|\partial_\lambda|0\rangle \mathrm e^{2i\eta^z
\int_{\lambda_i}^\lambda c(\lambda') d\lambda'}\right|^2.
\label{n_ex2}
\ee
If the integral over $\eta$ converges at large $\eta$ and the matrix element $\langle k|\partial_\lambda|0\rangle$ goes to a non-zero constant as $k\to 0$, then we  get the desired scaling. The second condition means that there is no kinematic suppression of the transitions to the low energy and momentum states. The absence of this suppression was implicitly assumed in the elementary derivation of Eq.~(\ref{d/z}). The first condition of convergence of the integral in Eq.~(\ref{n_ex2}) over $\eta$ implies that only low energy modes contribute to the total number of generated quasiparticles since $k\sim \eta \delta^{1/z}$. This happens only in the case where $d/z<2$, otherwise the quadratic scaling $n_{\rm ex}\sim \delta^2$, which we derived in the previous section, coming from excitations to all energy scales, is restored. Technically the crossover can be seen from the fact that the transition probability to the state with momentum $k$ at large $k\gg \delta^{1/z}$ corresponding to $\eta\gg 1$ scales as $1/\eta^{2z}$. This follows from combining Eqs.~(\ref{tr_prob}) and (\ref{disp}). Clearly then only if $d\leq 2z$ the integral over $\eta$ in Eq.~(\ref{n_ex2}) converges at large $\eta$. Therefore after the rescaling $k=\delta^{1/z}\eta$ the upper limit of integration over $\eta$ can be sent to infinity. Otherwise the quasiparticle excitations with large momenta independent of $\delta$ dominate $n_{\rm ex}$ and we obtain the quadratic scaling according to the general result (\ref{tr_prob}).

One can check that under the same conditions the excess energy or
heat~\cite{ap_heat} per unit volume in the system scales as:
\beq\label{heat}
&&Q\approx {1\over L^d}\sum_k
\varepsilon_k(\lambda_f)|\alpha_k(\lambda_f)|^2\nonumber=|\delta|^{d+z\over
z}c(\lambda_f) \nonumber\\ &&\times\int {d^d\eta\over
(2\pi)^d}\eta^z\left|\int\limits_{\lambda_i}^{\lambda_f} d\lambda\,\langle
\eta \delta^{1/z}|\partial_\lambda|0\rangle \mathrm e^{2i\eta^z
\int_{\lambda_i}^\lambda c(\lambda') d\lambda'}\right|^2.
\eeq
This scaling is now valid provided that $(d+z)/z\leq 2$. Otherwise the
energy absorbtion comes from the high energy states with $k\gg \delta^{1/z}$ and
in this case $Q\propto \delta^2$.

Since we assume that the quasiparticles are created independently in different
channels, i.e. the probability to create a pair of quasiparticles with momentum
$k$ is uncorrelated with the probability to create a pair of quasiparticles with
momentum $k'\neq k$, we can easily find the scaling of the (diagonal)
entropy density of the system~\cite{diag_ent}:
\be
S_d\approx-\frac{1}{2L^d}\sum_k |\alpha_k(\lambda_f)|^2\ln|\alpha_k(\lambda_f)|^2,
\label{ent2}
\ee
where the factor of $1/2$ comes from the fact that we need to account each state characterized by momenta $q$ and $-q$ once and sum only over half of the momentum modes to avoid over-counting. As before one can instead sum over all momentum modes but multiply the result by $1/2$. In the domain of validity of the adiabatic perturbation theory,
$|\alpha_k(\lambda_f)|^2\ll 1$, the expression for the entropy is very similar to the expression for the number of quasiparticles. The extra logarithm clearly does not affect the scaling with $\delta$ and we thus expect that
$S_d\propto |\delta|^{d/z}$ for $d/z\leq 2$.

{\em Example: harmonic chain.} Let us consider a specific example from
Ref.~\cite{pg_np}, where one slowly changes the mass of particles in a harmonic
chain. Specifically we consider the following Hamiltonian
\be\label{harm_chain}
\mathcal H=\sum_k {\rho_s k^2\over 2}|x_k|^2+{\kappa(t)\over 2}|p_k|^2,
\ee
where $x_k$ and $p_k$ are conjugate coordinate and momentum. We will assume that $\kappa$, playing the role of an inverse mass, linearly increases in time: $\kappa(t)=\kappa_i+\delta t$. For simplicity we will also assume that
$\kappa_f\gg\kappa_i$. In Ref.~\cite{pg_np} a more complicated situation with $\kappa$ dependent on $k$ was analyzed, however this $k$-dependence is only important if $\kappa_i\to 0$ corresponding to a singularity in the spectrum. In
this section we are interested in non-singular situations, therefore we assume that $\kappa_i$ is finite and that  an extra possible dependence of $\kappa$ on $k$ is irrelevant. The problem was analyzed both exactly and perturbatively in
Ref.~\cite{pg_np}. Here we briefly repeat the analysis and present additional results. Within the adiabatic perturbation theory changing $\kappa$ in time generates pairs of  quasiparticle excitations with opposite momenta. It is straightforward to check that the matrix element $\langle k,-k|\partial_\kappa|0\rangle=1/(4\sqrt{2}\kappa)$ is independent of $k$ satisfying the second requirement (see the sentence after Eq.~(\ref{n_ex2})) necessary to get the correct scaling of the excitations. Substituting this matrix element together with the dispersion $\epsilon_k=k\sqrt{\kappa\rho_s}$ into Eq.~(\ref{int_eq3}), we find that within the perturbation theory the probability to excite a pair of quasiparticles with momenta $k$ and $-k$ is
\beq
&&|\alpha_k|^2\approx {1\over 32}\left|\int_{\kappa_i}^{\kappa_f}
{d\kappa\over \kappa}\exp\left[{4i\over 3\delta}k\sqrt{\rho_s
}\kappa^{3/2}\right]\right|^2\nonumber\\ &&\approx {1\over
72}\left|\Gamma\left(0,-i{4\over 3}{k\kappa_i\sqrt{\rho_s\kappa_i}\over
\delta}\right)\right|^2,
\label{prob_chain_appr}
\eeq
where $\Gamma(0,x)$ is the incomplete $\Gamma$-function. In the equation above
we used the fact that $\kappa_f\gg \kappa_i$ so the upper limit in the  integration
over $\kappa$ can be effectively extended to infinity. The expression depends on the single dimensionless
parameter:
\begin{displaymath}
\xi_k={k\over k_\delta},
\end{displaymath}
where
\begin{displaymath}
k_\delta=\delta/\sqrt{\kappa_i^3\rho_s}.
\end{displaymath}

When $\xi_k\gg 1$, which corresponds to the high momentum modes, we have
\be
|\alpha_k|^2\approx {1\over 128\xi_k^2}={1\over 128}{k_\delta^2\over k^2},
\label{as1}
\ee
and in the opposite limit Eq.~(\ref{prob_chain_appr}) gives
\be
|\alpha_k|^2\approx {1\over 72} \left|\ln\xi_k\right|^2.
\label{as2}
\ee
Hence at high energies the transition probability is proportional to
$\delta^2$, as expected from the discussion in the previous section, while at
small momenta $k\to 0$ the transition probability diverges. This is of course
unphysical and indicates the breakdown of the perturbation theory. Note,
however, that because of the small prefactor $1/72$ this divergence occurs at
very small values of $\xi_k$.

This problem can also be solved exactly (see details in Ref.~\cite{pg_np}). The initial ground state
wave function can be written as a product of Gaussians:
\be\label{gauss}
\Psi_i(\{x_k\})=\prod_k {1\over
(2\pi\sigma^0_k(\kappa_i))^{1/4}}\exp\left[-{|x_k|^2\over
4\sigma^0_k(\kappa_i)}\right],
\ee
where
\begin{displaymath}
\sigma^0_k(\kappa)={1\over 2k}\sqrt{\kappa\over\rho_s}.
\end{displaymath}
One particularly useful property of Gaussian functions is that for arbitrary
time-dependence of $\kappa(t)$ (or $\rho_s(t)$) the wave function remains
Gaussian with $\sigma_k$ satisfying the first order differential equation:
\be
i{d\sigma_k(t)\over dt}=2\rho_s k^2\sigma_k^2(t)-{1\over 2}\kappa(t).
\label{eq:sig}
\ee
The solution of this equation which satisfies the proper initial condition can be
expressed through the Airy functions. It is convenient to introduce the effective
width of the wave function at time $t_f$:
\begin{displaymath}
\sigma_k^\star={1\over \Re(\sigma_k^{-1})}.
\end{displaymath}
It can be shown that~\cite{pg_np}:
\be
{\sigma_k^\star\over \sigma_k^0}={1+|f_k|^2\over 2\Im f_k},
\ee
where
\be\label{fun_f}
f_k=-{\sqrt[3]{\xi_k}\,{\rm Bi}(-\xi_k^{2/3})-i{\rm Bi'}(-\xi_k^{2/3})\over
\sqrt[3]{\xi_k}\,{\rm Ai}(-\xi_k^{2/3})-i{\rm Ai'}(-\xi_k^{2/3}) }.
\ee
As it occurs in the perturbative treatment, the solution is expressed entirely
through the single parameter $\xi_k$. It is easy to show that the average number
of excited particles with momenta $k$ and $-k$ in the Gaussian state characterized by
the width $\sigma_k^\star$ is
\be
n_k={1\over 2}\left[{\sigma_k^\star\over \sigma_k^0}-1\right].
\ee
This expression has the following asymptotics:
 for $\xi_k\gg1$
\be
n_k\approx {1\over 64\xi_k^2}
\ee
and for $\xi_k\ll 1$
\be
n_k\approx {\pi \over 3^{2/3}\Gamma^2(1/3)}{1\over \sqrt[3]{\xi_k}}.
\ee
For large momenta the exact asymptotic clearly
coincides with the approximate one~(\ref{as1}) (noting that
$n_k=2|\alpha_k|^2$), while at low energies as we anticipated the adiabatic
perturbation theory fails. The perturbative result clearly underestimates the
number of excitations. This fact is hardly surprising because the adiabatic
perturbation theory neglects the tendency of bosonic excitations to bunch together,
leading to the enhancement of the transitions. We can anticipate that the
result will be opposite in the fermionic case and in the next section we will
see that this is indeed the case.

{\em Density of quasiparticles.} The density of quasiparticles created
during the process can be obtained by summing $n_k$ over all momenta:
\be
n_{\rm ex}={1\over 2}\int {d^dk\over (2\pi)^d} n_k.
\label{n_ex}
\ee
We remind that the factor of $1/2$ is inserted to avoid double counting of pairs with momenta $k$ and $-k$. In low dimensions $d\leq 2$ this sum is dominated by low momenta $k\sim k_\delta$ and the upper limit can be send to $\infty$:
\be
n_{\rm ex}\approx {k_\delta^d\over 2}\int {d^d\xi\over (2\pi)^d}n_k(\xi).
\label{nex_chain}
\ee
This gives the correct scaling $n_{\rm ex}\sim \delta^{d/z}$ anticipated
from the general argument since $z=1$. On the other hand above two dimensions the
integral over $k$ is dominated by high momenta close to the high energy
cutoff and we have
\be
n_{\rm ex}\approx {1\over 128}k_\delta^2\int {d^dk\over (2\pi)^d} {1\over k^2}.
\ee
This result again confirms our expectations that the number of excited
quasiparticles scales as $\delta^2$ in high dimensions. Note that when $d=2$
the integral above is still valid but it should be cutoff at small $k\sim
k_\delta$, leading to the additional log-dependence of $n_{\rm ex}$ on
$\delta$. We point out that the perturbative analysis predicts a very
similar qualitative picture: it correctly predicts the density of excitations in high
dimensions $d\geq 2$ and gives the correct scaling in low dimensions $d<2$.
However, in the latter case the perturbative analysis gives a wrong prefactor. We
note that there are situations when the adiabatic perturbation theory can fail completely
predicting even incorrect scaling~\cite{pg_np}. This can happen, for example,
if the initial coupling $\kappa_0$ is very small. Then the integral
(\ref{n_ex}) can become infrared divergent at small momenta and should be
cutoff by the inverse system size or another large spatial scale (e.g. the mean free
path).

The smallest (and the only) physical dimension where the scaling
$n_{\rm ex}\propto |\delta|^{d/z}$ is valid in our situation is $d=1$. Accurate evaluation of the quasiparticle density in this case gives
\be
n_{\rm ex}\approx 0.0115 k_\delta.
\ee
For completeness we also quote the perturbative result obtained by integrating
Eq.~(\ref{prob_chain_appr}):
\be
n_{\rm ex}^{\rm pert}\approx 0.0104 k_\delta.
\ee
Clearly the difference between the exact and perturbative results is very
minor in this case.

In two dimensions both perturbative and exact treatments give the same result
\be
n_{\rm ex}\approx {1\over 256\pi}k_\delta^2\ln\left({\Lambda\over
k_\delta}\right),
\ee
where $\Lambda$ is the short distance cutoff. At higher dimensions, as we
mentioned above, results of perturbative and exact treatments are identical
and non-universal:
\be
n_{\rm ex}\approx C k_\delta^2\Lambda^{d-2}.
\ee
It is interesting to note that adiabatic conditions are determined now by the
high momentum cutoff: $\delta\sim 1/\Lambda^{d/2-1}$. This sensitivity to $\Lambda$ might be
an artifact of a sudden change in the rate of change of the parameter $\kappa$ resulting in the infinite acceleration $\ddot\lambda$ at initial (and final) times. However, analyzing this issue in detail is beyond the scope of this work.

{\em Heat (excess energy)}. The quasiparticle density is not always an easily
detectable quantity, especially if the system is interacting and the number of
quasiparticles is not conserved. A more physical quantity is the energy
change in the system during the process, which is equal to the external work
required to change the coupling $\kappa$. This energy consists of two parts:
adiabatic, which is related to the dependence of the ground state energy on
$\kappa$, and heat, i.e. the additional energy pumped into the system due to
excitations of higher energy levels. The first adiabatic contribution depends only
on the initial and final couplings $\kappa_i$ and $\kappa_f$ but not on the
details of the process. On the contrary the heat (per unit volume) $Q$ is directly related to
the rate $\delta$. The microscopic expression for $Q$ can be obtained by a
simple generalization of Eq.~(\ref{n_ex}), as shown in Eq.~(\ref{heat}):
\be
Q={\sqrt{\kappa_f\rho_s}\over 2}\int {d^d k\over (2\pi)^d}\, k\, n_k.
\ee
This integral converges at large $k$ only for $d<1$. Therefore in all physical dimensions it is dominated by the
high energy asymptotic~(\ref{as1}) so $Q\propto \delta^2$. In one dimension there is an extra logarithmic correction
\be
Q_{d=1}\approx {k_\delta^2\over
256\pi}\sqrt{\kappa_f\rho_s}\ln\left({\Lambda\over k_\delta}\right).
\ee

{\em Entropy}. Finally let us evaluate the generated diagonal entropy in the
system. The latter is formally defined as (see Ref.~\cite{diag_ent})
\be
S_d=-{1\over L^d}\sum_n \rho_{nn}\log \rho_{nn},
\label{ent1}
\ee
where $\rho_{nn}$ are the diagonal elements of the density matrix in the
eigenbasis of the (final) Hamiltonian. At finite temperatures it is this entropy which is connected to heat via the standard thermodynamic relation $\Delta Q=T\Delta S_d$. However, at zero initial temperature this relation breaks down and one should analyze $S_d\equiv \Delta S_d$ separately.

Since in our problem different momentum states are decoupled the d-entropy is additive:
\be\label{ent3}
S_d={1\over 2L^d}\sum_k s_k={1\over 2}\int {d^dk\over (2\pi)^d} s_k,
\ee
where the factor of $1/2$ is again present in order to avoid double counting of quasiparticle excitations created in pairs. Within the adiabatic perturbation theory only one pair of quasiparticles can be excited. Therefore $s_k^{\rm pert}\approx -|\alpha_k|^2\ln |\alpha_k|^2$. Note that at large $k$ the entropy per mode falls off with $k$ essentially in the same manner as $|\alpha_k|^2$ so we conclude that the entropy is dominated by small energies below two dimensions. Then we find
\be
S_d^{\rm pert}\approx -{1\over 2} k_\delta^d\int {d^d\xi\over (2\pi)^d} |\alpha(\xi)|^2 \ln|\alpha(\xi)|^2.
\label{sd1}
\ee
In dimensions higher than two the entropy is dominated by high momenta so
\be
S_d^{\rm pert}\approx -{1\over 256}k_\delta^2\int {d^dk\over (2\pi)^d} {1\over k^2}\ln\left({k_\delta^2\over 128k^2}\right).
\label{sd2}
\ee
In one dimension Eq.~(\ref{sd1}) gives
\be
S_{d=1}^{\rm pert}\approx 0.02 k_\delta.
\ee

To calculate the exact value of the entropy density one needs to project the
Gaussian wave function describing each momentum mode to the eigenbasis and
calculate the sum~(\ref{ent1}). There is, however, a simple shortcut allowing
to instantly write the answer. One can easily check that the probabilities to occupy different eigenstates are identical to those of the equilibrium thermal ensemble. Therefore the entropy per mode can be expressed through the average number of excited quasiparticle pairs $n_k/2$ as
\be
s_k=-(n_k/2)\ln (n_k/2) + (1+n_k/2)\ln (1+n_k/2).
\ee
Using explicit solution for $n_k$ we find
\be
S_{d=1}\approx 0.026 k_\delta.
\ee
The result is again quite close to the perturbative one.

In two dimensions the entropy density is also readily
available from the expressions above:
\be
S_{d=2}\approx {k_\delta^2\over 512\pi}\left[\ln(128 e) \ln\left({\Lambda\over
k_\delta}\right)+\ln^2\left({\Lambda\over k_\delta}\right)\right].
\ee

\section{Adiabatic dynamics near a quantum critical point}

Let us apply the analysis of the previous section to the case of crossing a
quantum critical point. As before we consider the situation in which the
system is prepared in the ground state, characterized by some initial coupling
$\lambda_i$,  then this coupling is linearly tuned in time until a finite
value $\lambda_f$. We assume that the system undergoes a second order quantum phase transition at $\lambda=0$. We
differentiate the two situations: (i) when $\lambda_i$ is finite and negative
and $\lambda_f$ is finite and positive and (ii) when either $\lambda_i=0$ and
$\lambda_f$ is large and positive, or $\lambda_f=0$ and $\lambda_i$ is large
and negative. As in the previous section we will first give a general discussion
closely following Ref.~\cite{ap_adiabatic} and then analyze specific examples.

\subsection{Scaling analysis}\label{gen_scaling}

The non-adiabatic effects are especially pronounced in the vicinity of a quantum critical point (QCP), where one can expect universality in the transition rates. Near a QCP for $\lambda\neq 0$ the system develops a characteristic energy scale $\Delta$, vanishing at the critical point as $\Delta \sim |\lambda|^{z \nu}$, where $z$ is the dynamical exponent, and $\nu$ is the critical exponent of the correlation length, $\xi\sim
|\lambda|^{-\nu}$~\cite{subir}. This energy scale can be either a gap or some crossover scale at which the energy spectrum qualitatively changes.

 As before, let us perform the scaling analysis of Eq.~(\ref{int_eq3}) assuming that  the quasiparticles are created in pairs of opposite momenta. First we rewrite the dynamical phase factor entering the expression for the transition amplitude~(\ref{int_eq3}) as
\be\label{phase}
\Theta_k(\lambda)-\Theta_0(\lambda)={1\over\delta}\int_{\lambda_i}^\lambda
d\lambda' (\varepsilon_k(\lambda')-\varepsilon_0(\lambda')).
\ee
Near the QCP the quasiparticle energy can be rewritten using the scaling function $F$ as:
\be
\varepsilon_k(\lambda)-\varepsilon_0(\lambda)=\lambda^{z\nu}
F(k/\lambda^{\nu}).
\label{en_sc}
\ee
For $x \gg 1$ the function $F(x)$ should have asymptotic $F(x)\propto x^z$, reflecting the
fact that at large momenta the energy spectrum should be insensitive to $\lambda$.  At small $x$, corresponding to small $k$, the asymptotical behavior of the scaling function $F(x)$ depends on whether the system away from the
singularity is gapped, then $F(x)\to$const at $x\to 0$, or gapless, then $F(x)\propto x^\alpha$ with some positive exponent $\alpha$. The scaling~(\ref{en_sc}), inserted in (\ref{phase}), suggests the change of variables
\be\label{scaling}
\lambda=\xi\,\delta^{1\over z\nu+1},\quad k=\eta\, \delta^{\nu\over z\nu+1}.
\ee
In addition  to analyze Eq.~(\ref{int_eq3}) we adopt the scaling ansatz for the matrix element
\be
\langle k|\partial_{\lambda}|0\rangle=-{\langle k|V|0\rangle\over \varepsilon_k(\lambda)-\varepsilon_0(\lambda)}= {1\over \lambda}G(k/\lambda^{\nu}),
\label{scaling_matr_el}
\ee
where $G(x)$ is another scaling function. This scaling dimension of the matrix element, which is the same as the engineering dimension $1/\lambda$, follows from the fact that the ratio of the two energies $\langle k|\lambda V|0\rangle$ and $\varepsilon_k(\lambda)-\varepsilon_0(\lambda)$ should be a dimensionless quantity. At large momenta $k\gg \lambda^\nu$ this matrix element should be independent on $\lambda$ so $G(x)\propto x^{-1/\nu}$ at $x\gg 1$. This statement must be true as long as the matrix element $\langle k|V|0\rangle$ is non-zero at the critical point, which is typically the case. In the opposite limit $x\ll 1$ we expect that $G(x)\propto x^\beta$, where $\beta$ is some non-negative number.

These two scaling assumptions allow one to make some general conclusions on the behavior of the density of
quasiparticles and other thermodynamic quantities with the quench rate for the adiabatic passage through a
quantum critical point. Thus
\be
n_{\rm ex}\sim \int {d^dk \over (2\pi)^d }|\alpha_k|^2=|\delta|^{d\nu\over
z\nu+1} \int {d^d\eta\over (2\pi)^d} |\alpha(\eta)|^2,
\label{n_ex1}
\ee
where, after the rescaling (\ref{scaling}),
\be
\alpha(\eta)=\int_{\xi_i}^{\xi_f} d\xi {1\over
\xi}G\left({\eta\over\xi^{\nu}}\right)\exp\left[i\int_{\xi_i}^\xi d\xi_1
\xi_1^{z\nu}F(\eta/\xi_1^{\nu})\right].
\label{alpha}
\ee
Note that if $\lambda_i<0$ and $\lambda_f>0$ then $\xi_i$ is large and negative and $\xi_f$ is large and positive. If we start (end) exactly at the critical point then $\xi_i=0$ ($\xi_f=0$). Note that the integral over $\xi$ is always convergent because: at large $\xi$ we are dealing with a fast oscillating function, and at $\xi\sim 0$ there are no singularities because of the scaling properties of $G(x)$. The only issue which has to be checked carefully is the convergence of the integral over $\eta$ at large $\eta$. If this integral converges then Eq.~(\ref{n_ex1}) gives the correct scaling relation for the density of quasiparticles with the rate $\delta$ found in earlier works~\cite{ap_adiabatic, zurek_adiabatic}. If the integral does not converge then the density of created defects is dominated by high energies and from general arguments of Sec.~\ref{Sec:pert} we expect that $n_{\rm ex}\propto \delta^2$. As it is evident from Eq.~(\ref{n_ex1}), the crossover between these two regimes happens when  $d=2z+2/\nu$~\endnote{We note that there is a small error in Ref.~\cite{ap_adiabatic}, which gives a different expression for $d_c$.}. To see how this crossover emerges from Eq.~(\ref{n_ex1}) we analyze the asymptotical behavior of $\alpha(\eta)$ at $\eta\gg \xi^{\nu}$. In this limit the integral over $\xi$  can be evaluated in accord to the discussion given in Sec.~\ref{Sec:pert} because the exponent in Eq.~(\ref{alpha}) is a rapidly oscillating function of $\xi$. Using the explicit asymptotics of the scaling functions $F(x)$ and $G(x)$ at large $x$ we find
\be
\alpha(\eta)\propto {1\over \eta^{z+{1/\nu}}}.
\ee
Then the integral over $\eta$ in Eq.~(\ref{n_ex1}) converges at large $\eta$
 precisely when $d\leq 2z+2/\nu$. Instead when the integral does not converges,
  the scaling $n_{\rm
 ex}\propto |\delta|^{d\nu/(z\nu+1)}$ breaks down  and  we get $n_{\rm ex}\propto \delta^2$. As in the
 previous section we can expect logarithmic corrections at the crossover between these two scaling behaviors. 

One can similarly analyze the dependence of the heat $Q$ and the diagonal entropy $S_d$ on
the rate $\delta$. The entropy has the same scaling
as the number of quasiparticles. We note that in general the heat is universal
only if the process ends at the critical point $\lambda_f=0$, where
$\epsilon_k\propto k^z$. Then it is easy to see that
\be
Q\propto |\delta|^{(d+z)\nu\over z\nu +1}.
\label{Q_ex}
\ee
 This scaling is valid for $d\leq z+2/\nu$, while it becomes quadratic when $d > z+2/\nu$. If the final coupling $\lambda_f$ is away from the critical point then the $\delta$-dependence of $Q$ becomes sensitive to the behavior of the spectrum. Thus if there is a gap in the spectrum, $Q$ has the same scaling as $n_{\rm ex}$ and $S_d$,
since each excitation roughly carries the same energy equal to the gap. If the spectrum at $\lambda_f$ is gapless then the scaling (\ref{Q_ex}) remains valid.

\subsection{Examples}

Here we will consider several specific models illustrating the general predictions
above. In particular we will analyze the transverse field Ising model, which
serves as a canonical example of quantum phase transitions~\cite{subir} and
which was used as an original example where the general scaling~(\ref{n_ex1})
was tested~\cite{ap_adiabatic, zurek_adiabatic, jacek}. The transverse field
Ising model also maps to the problem of loading one-dimensional hard-core
bosons or non-interacting fermions into a commensurate optical lattice
potential~\cite{claudia} and describes the so-called Toulouse point in the
sine-Gordon model, where this model can be mapped to free spinless fermions~\cite{giamarchi_book}.

\subsubsection{Transverse field Ising and related models}\label{Ising}

The transverse-field Ising model is described by the following Hamiltonian:
\be
\mathcal H_{\rm I}=-\sum_j \left[g(t)\sigma_j^x+\sigma^z_j\sigma^z_{j+1}\right].
\label{ham_quadratic}
\ee
For simplicity we will focus only on the domain of non-negative values of the transverse field $g$. This model undergoes a quantum phase transition at $g=1$~\cite{subir} with $g > 1$ corresponding to the transversely magnetized phase and $g < 1$ corresponding to the phase with  longitudinal magnetization. It is thus convenient to use $\lambda(t)=g(t)-1$  as the tuning parameter. Under the Jordan-Wigner transformation:
\be
\sigma^z_j=-(c_j+c_j^\dagger)\prod_{i<j} (1-2c_i^\dagger c_i),\; \sigma^x_j=1-2c_j^\dagger c_j
\ee
the Hamiltonian assumes the quadratic form:
\be
\mathcal H_{\rm I}=-\sum_j c_j^\dagger c_{j+1}+c_{j+1}^\dagger c_j+c_j^\dagger c_{j+1}^\dagger + c_{j+1} c_j-2g(t)c_j^\dagger c_j
\ee
and can be diagonalized using the Bogoliubov transformation in the momentum space:
\be
c_k=\gamma_k \cos(\theta_k/2) +i\sin(\theta_k/2)\gamma_{-k}^\dagger,
\ee
where
\be
\tan\theta_k={\sin(k)\over \cos(k)-g(t)}.
\ee
After this transformation the Hamiltonian becomes
\be
\mathcal H_I=\sum_k \varepsilon_k\gamma_k^\dagger\gamma_k,
\ee
where
\be
\varepsilon_k=2\sqrt{1+g^2-2g\cos(k)}.
\ee
The ground state of this Hamiltonian factorizes into the product:
\be
|\Omega_0\rangle=\prod_k \left(\cos(\theta_k/2)+i \sin(\theta_k/2)c_k^\dagger c_{-k}^\dagger\right)|0\rangle.
\label{gs}
\ee
The excited states can be obtained by applying various combinations of operators $\gamma_k^\dagger$ to the ground state above. However, because of the momentum conservation, only the excited states obtained by acting on the ground state by the products $\gamma_k^\dagger\gamma_{-k}^\dagger$ are relevant. Because the excitations to different momentum states are independent, the problem effectively splits into a sum of independent Landau-Zener problems~\cite{jacek} and can be exactly solved. In the case when $g_i$ and $g_f$ lie on the opposite sides of the quantum critical point the transition probability in the slow limit is approximately given by~\cite{jacek}
\be
p_k\approx\exp\left[-{2\pi k^2\over \delta}\right].
\ee
The density of the excited quasiparticles is then
\be
n_{\rm ex}={1\over 2\pi}\int_{-\infty}^{\infty} p_k dk\approx {\sqrt{\delta}\over 2\pi\sqrt{2}}\approx 0.11\sqrt{\delta}.
\ee
Similarly one can find the entropy density generated during the process
\beq
&&S_d\approx -{1\over 4\pi}\int_{-\infty}^{\infty} dk\left[p_k\ln p_k+(1-p_k)\ln(1-p_k)\right]\nonumber\\
&&~~~\approx 0.052\sqrt{\delta}.
\label{sd_ising}
\eeq
Both the expressions for $n_{\rm ex}$ and $S_d$ agree with the general scaling law (\ref{n_ex1}) with $d=z=\nu=1$. Since in the final state all the excitations are gapped the heat in this case is approximately equal to the number of excited quasiparticles multiplied by the gap in the final state and thus has the same scaling as $n_{\rm ex}$.

The same problem can be also solved using the adiabatic perturbation theory. Let us note that as it follows from Eq.~(\ref{gs}) the transition matrix element reads~\cite{ap_adiabatic}
\be
\langle k,-k|\partial_\lambda|0\rangle=\langle k,-k|\partial_g|0\rangle={i\over 2}{\sin k\over 1+g^2-2g\cos(k)}.
\ee
In the limit of small $\delta$ only the transitions happening in the vicinity of the critical point contribute to $n_{\rm ex}$. In this case we have
\be\label{el_ising}
\langle k,-k|\partial_\lambda|0\rangle\approx{i\over 2}{k\over \lambda^2+k^2},
\ee
which clearly satisfies the scaling (\ref{scaling_matr_el}). Under the same approximation we can use that $\varepsilon_k\approx 2\sqrt{k^2+\lambda^2}$, which in turn satisfies the energy scaling~(\ref{en_sc}). Substituting these expansions into Eq.~(\ref{int_eq3}) and extending the limits of integration over $\lambda$ to $(-\infty,\infty)$ we find that
\be\label{alfa_ising}
\alpha_k\approx -{i\over 2}\int_{-\infty}^{\infty} d\lambda{k\over \lambda^2+k^2} \exp\left[{4 i\over \delta}\int_0^\lambda d\lambda'\sqrt{k^2+\lambda'^2}\right],
\ee
where the additional factor of two in the exponent comes from the fact that two quasiparticles are created during each transition. It is now straightforward to evaluate the perturbative expression for $n_{\rm ex}^{\rm pert}$:
\be
n_{\rm ex}^{\rm pert}\approx\int_{-\infty}^{\infty} {dk\over 2\pi} |\alpha_k|^2 \approx 0.21 \sqrt{\delta}.
\ee
Note that unlike  the bosonic case discussed in the previous section, the adiabatic perturbation theory now overestimates the number of created quasiparticles. This happens because this perturbation theory does not take into account the Pauli blocking which prevents more than one pair of quasiparticles with momenta $k,-k$ to be excited. Technically the perturbative transition probability $|\alpha_k|^2$ can exceed unity. This prevents  from computing the generated entropy using Eq.~(\ref{ent2}) because there will be spurious negative contributions. To get a sensible expression one needs to integrate only over the momenta satisfying $|\alpha_k|^2\leq 1$, and this results in:
\beq
&&S_d^{\rm pert}\approx -{1\over 2\pi}\int\limits_{k_{min}}^\infty dk |\alpha_k|^2\ln |\alpha_k|^2\approx 0.022\sqrt{\delta}.
\eeq
The perturbative argument can be somewhat improved explicitly using the fact that the quasiparticles are fermions and adding an additional contribution coming from $(1-|\alpha_k|^2)\ln(1-|\alpha_k|^2)$. In this case $S_d^{\rm pert}\approx 0.038\sqrt{\delta}$, which is closer to the exact result.

In a similar spirit we can consider the situation when either $\lambda_i=0$ or $\lambda_f=0$, i.e. when the initial or final state exactly corresponds to the quantum critical point. This situation requires fine tuning from the point of view of crossing a quantum phase transition. However, it naturally appears in other  contexts. For example, the problem of loading hard core bosons into a commensurate periodic potential exactly describes this situation~\cite{claudia}.  Since we consider $\lambda \in (0,\infty)$ then Eq.~(\ref{alfa_ising}) becomes:
\be
\alpha_k\approx -{i\over 2}\int_0^{\infty} d\lambda{k\over \lambda^2+k^2} \exp\left[{4 i\over \delta}\int_0^\lambda d\lambda'\sqrt{k^2+\lambda'^2}\right],
\ee
then it is easy to see that $n_{\rm ex}^{\rm pert}(0,\infty)=\frac{1}{4}n_{\rm ex}^{\rm pert}(-\infty,\infty)$, therefore the scaling of the density of quasiparticles (and entropy) remains the same $n_{\rm ex}\propto \sqrt{\delta}$. This problem can be also solved exactly and the results remain very close to the perturbative case. We will postpone a careful analysis of this problem to the next section.

\subsubsection{sine-Gordon model: Toulouse point and  limit of free massive bosons}
\label{sec:sg}

A convenient playground to test the general scaling laws presented in Sec.~\ref{gen_scaling} is the sine-Gordon model (SG) -- a one dimensional model described by the Hamiltonian:
\beq\label{SG}
\HH = \frac{1}{2}\int d\,x\, \lb \Pi(x)^2 +(\partial_x \phi)^2-4\lambda \cos(\beta \, \phi)\rb.
\eeq
Here $\Pi(x)$ and $\phi(x)$ are conjugate fields, $\lambda$ is the tuning parameter and $\beta=2 \sqrt{\pi K_{\rm SG}}$ is a constant. From the renormalization group (RG) analysis it is known that the cosine-term  is a relevant perturbation to the quadratic model only if $0\leq K_{\rm SG} <2$, and therefore the system is gapped at any finite $\lambda$~\cite{giamarchi_book},  while for $K_{\rm SG}>2$ the system remains gapless. Turning on the interaction in the regime $K_{\rm SG}<2$ is akin starting at the critical point and driving the system into the new gapped phase. Hence the scaling of the density of excitations and other quantities should be described by the critical exponents in accord to Eq.~(\ref{n_ex1}) as it was indeed shown in Ref.~\cite{claudia}.  The spectrum of the SG model consists of solitons and antisolitons  for $1\leq K_{SG}\le 2$ and in addition to those for  $K_{SG}<1$ there are also breathers excitations. The point $K_{SG}=1$ is the Toulouse point described by free fermions and in the limit $K_{SG}\to 0$ the SG model effectively describes a system of free massive bosons with $B_1$ breathers being the only surviving excitations~(see e.g. Ref.~\cite{vova_lr}). The energy spectrum for each momentum $k$ is of the form:
\be\label{SG_energy}
\epsilon_k=\sqrt{k^2 +m_s^2},
\ee
where $m_s$ is the soliton (or breather) mass, that scales with the external parameter as $m_s\sim \lambda^{1/(2-K_{SG})}$. Therefore it is evident  how this model fulfills the assumption (\ref{en_sc}) with the critical exponents $z=1$ and $\nu=1/(2-K_{SG})$.

The SG model gives the correct low energy description of: (i) interacting bosons in a commensurate periodic potential and (ii) two one-dimensional condensates (Luttinger liquids) coupled by a tunneling term~\cite{claudia}. In the former case increasing $\lambda$ in time corresponds to  loading bosons into an optical lattice and in the latter case increasing $\lambda$ describes turning on the tunneling coupling. We note that $\phi$ describes the density modulation in the first situation and the relative-phase between the two  superfluids in the second. The SG model also naturally appears in several other one-dimensional systems~\cite{giamarchi_book}. To simplify the analysis we will consider only the situation where either $\lambda_i=0$ and $\lambda_f$ large or vice versa. For a generic value of $K_{SG}$ the problem can only be solved perturbatively~\cite{claudia} and the corresponding discussion is beyond the scope of this work. Here we will discuss only two specific solvable limits $K_{SG}\ll 1$ and $K_{SG}=1$.

\paragraph{$K_{SG}\ll 1$: free bosons.}
This limit can be naturally realized in the situation of merging two weakly interacting one-dimensional condensates. Then $K_{SG}=1/(2K)$, where $K\gg 1$ is the Luttinger liquid parameter describing the individual condensates~\cite{claudia}.
Since $K_{SG}$ and hence $\beta$ in the Hamiltonian (\ref{SG}) is small we can expand the cosine term and get a quadratic Hamiltonian that in the Fourier space has the form:
\be\label{SGqua}
\HH = {1\over 2}\sum_k  | \Pi_k|^2 +\kappa_k(t)|\phi_k|^2,
\ee
with $\kappa_k(t)=k^2+2\lambda(t)\beta^2$, where as before we assume that $\lambda(t)=\delta t$. We see that in this case the problem is equivalent to the harmonic chain (\ref{harm_chain}) already considered in Section \ref{gapless}. The only difference is in the excitations spectrum $\epsilon_k=\sqrt{\kappa_k(t)}$ which is now gapped:
\be
\epsilon_k=\sqrt{k^2 + c \lambda},
\ee
with $c=2\beta^2$. The perturbative and exact solutions follow the similar steps showed in Sec. \ref{gapless}. In the perturbative case we find
\beq
&&|\alpha_k|^2\approx {1\over 32}\left|\int_{\kappa_i}^{\kappa_f} {d\kappa\over \kappa}\exp\left[{4i\over 3 c \delta}\kappa^{3/2}\right]\right|^2\nonumber\\
&&\approx {1\over 72}\left|\Gamma\left(0,-i{4\over 3}{k ^3\over c \delta}\right)\right|^2,
\eeq
where we used the fact that $\kappa_i=k^2$. It is convenient to introduce the rescaled momentum
\be
\zeta_k=\frac{k}{k_\delta},
\ee
where
\be
k_\delta=(c \delta)^{1/3}.
\ee
Then all the results can be found from those of  the harmonic chain (cf. Eq.~(\ref{prob_chain_appr})), performing the mapping
\begin{displaymath}
\xi_k \longleftrightarrow \zeta_k^3.
\end{displaymath}
In the limit  $\zeta_k\gg 1$, corresponding to the high momentum modes, the  transition probability becomes
\be
|\alpha_k|^2\approx {1\over 128\zeta_k^6}={1\over 128}{(c \delta)^2\over k^6}
\ee
and in the opposite limit
\be
|\alpha_k|^2\approx {1\over 72} \left|\ln\zeta_k^3\right|^2.
\ee
For the exact solution of the problem we again follow the Gaussian functions ansatz as in Eq.~(\ref{gauss}), with initial value
\begin{displaymath}
\sigma^0_k(\kappa)=\frac{\sqrt{\kappa_k(0)}}{2}
\end{displaymath}
and time dependence satisfying  the equation:
\be
i{d\sigma_k(t)\over dt}=2 \sigma_k^2(t)-{1\over 2}\kappa(t).
\label{eq:sig1}
\ee
The solution of this equation is analogous to the one of Eq.~(\ref{eq:sig}) with the only difference that the function $f_k$ (cf. Eq.~(\ref{fun_f})) now becomes
\be
f_k=-{\zeta_k\,{\rm Bi}(-\zeta_k^2)-i{\rm Bi'}(-\zeta_k^2)\over \zeta_k\,{\rm Ai}(-\zeta_k^2)-i{\rm Ai'}(-\zeta_k^2)},
\ee
which gives the following asymptotics for the average number of excited particle pairs with momenta $k$ and $-k$:
\be
n_k\approx {1\over 64\zeta_k^6}={1\over 64}{(c\delta)^2\over k^6}
\ee
for $\zeta_k\gg 1$ and
\be
n_k\approx {\pi \over 3^{2/3}\Gamma^2(1/3)}{1\over \zeta_k}={\pi \over 3^{2/3}\Gamma^2(1/3)}{(c \delta)^{1/3}\over k}
\label{nk2}
\ee
in the opposite limit.

{\em Density of quasiparticles.} We note that the exact result gives weak logarithmic divergence of the density of quasiparticles with the system size coming from the $1/k$ dependence of $n_k$ at small $k$ (see Eq.~(\ref{nk2})):
\be
n^{\rm exact}_{\rm ex}\approx {(c \delta)^{1/3}   \log(k_\delta L)\over 2 \sqrt[3]{3^2}\Gamma^2(1/3)}\approx 0.033 (c \delta)^{1/3}\log(k_\delta L),
\ee
while the perturbative solution gives:
\be
n^{\rm pert}_{\rm ex}\approx 0.068 (c \delta)^{1/3}.
\ee
As previously found, the perturbative result under-estimates the number of excitated quasiparticles, since it does not take into account the bosonic enhancement of the transitions. Emergence of the length dependence in the expression for $n_{\rm ex}$ indicates the breakdown of the adiabatic perturbation approach and corresponds to a different (non-adiabatic) response of the system according to the classification of Ref.~\cite{pg_np}. Physically this divergence comes from the effect of bunching of bosons at small momenta and overpopulation of low momentum modes, which can not be captured in the lowest order of the adiabatic perturbation theory. A similar analysis can be also performed for the heat and the entropy. In the situation in which the system starts at the critical point ($\lambda_i=0$) and $\lambda_f$ is finite, both the entropy and the heat have a very similar behavior as $n_{\rm ex}$, i.e. proportional to $\delta^{1/3}$ and showing a weak logarithmic divergence with the system size. In the opposite case where $\lambda_i$ is finite but $\lambda_f=0$ the expressions for the density of excitations and entropy do not change, while the expression for the heat becomes different because each mode now carries energy $\epsilon_k\sim k$ proportional to the momentum. This removes the logarithmic divergence and both perturbative and exact results give $Q\sim k_\delta^2\sim (c\delta)^{2/3}$

\paragraph{$K_{SG}=1$: Tonks-Girardeau gas.}\label{TGgas}
In the limit of $K_{SG}=1$ the repulsive interaction between  bosons is infinitely strong, therefore the particles behave as impenetrable spheres (hard-core bosons). It is well known that in this limit the system known as the Tonks-Girardeau gas~\cite{girardeaus} can be mapped into an equivalent system of free spinless fermions (the corresponding limit of the SG model describes the so called Toulouse point). Therefore the dynamical problem  of loading hard-core bosons into a commensurate optical lattice  can instead be approached with the much simpler analysis of free fermions in a periodic potential.

\begin{figure}[ht]
\includegraphics[width=3.2in]{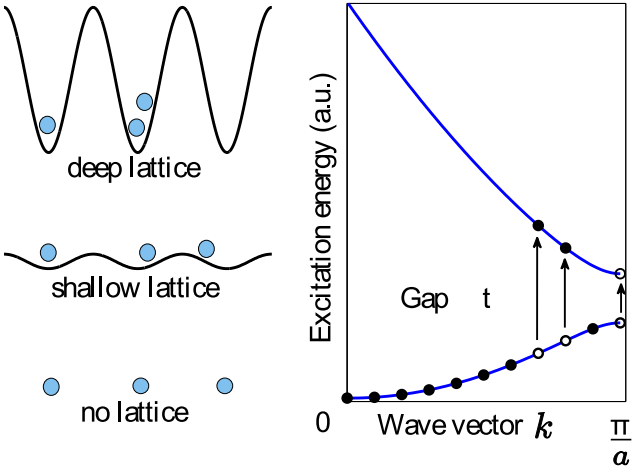}
\caption{TG-gas: the loading into a commensurate lattice problem can be mapped into a two-level system composed of lower filled band and upper empty band, the excitations are the particles hopping on the upper band (figure taken from ~\cite{claudia}).}
\label{band_gap}
\end{figure}

To understand the dynamics in this case we need to solve the Schr\"odinger equation of free fermions in a periodic potential with time dependent amplitude $V(x,t)=V(t) \cos(2k_f x)$, where $k_f=\pi/a$ is the Fermi momentum and $a$ is the lattice spacing. The potential $V(t)$ is related to the coupling $\lambda(t)$ in the SG Hamiltonian (\ref{SG}) according to $V(t)n=4\lambda(t)$, where $n=k_f/\pi$ is the electron density~\cite{buchler,claudia}. We assume $V(t)=\delta_V t$, therefore the rate $\delta_V$ is  related to the rate $\delta$ by simple rescaling: $\delta_V=4\delta/n$.
We restrict the analysis to the two lowest bands of the Brillouin zone (Fig. \ref{band_gap}) and we linearize the spectrum close to the Fermi momentum. These two approximations, justified for analyzing the low energy excitations that we are interested in, make the fermion problem equivalent to the SG problem. It follows that for each momentum $k$ the problem is described by a Landau-Zener Hamiltonian
\be
\HH_k=\left[\begin{matrix} V(t)/ 2 & \Delta_k \\ \Delta_k & -V (t)/2 \end{matrix}\right],
\ee
where  $\Delta_k  =  (\epsilon_k-\epsilon_{k-2 k_f})/2$ is half the energy difference between the 1st and the 2nd band. In the linearized approximation (under which the mapping to SG model is valid) we have $\epsilon_k\approx v_f (k-k_f)$, where $v_f$ is the Fermi velocity. Choosing the units where $v_f=1$ makes the free fermion problem identical to the SG problem with the Hamiltonian~(\ref{SG}).

The problem of turning on the potential amplitude from zero maps then to a direct sum of \textit{half} LZ-problem ($t\in [0,+\infty)$) (see Appendix~\ref{LZ_app}) and can be solved exactly (see also Ref.~\cite{damski_zurek}). Then the probability of exciting a particle from the lower to the upper band during the loading process is:

\beq
&&p_{\rm ex}(\tau_k)=1-2\frac{ e^{-\pi \tau_k/8}}{\pi
  \tau_k}\sinh\left(\frac{\pi \tau_k}{4}\right) \nonumber\\
&&\times\left|\Gamma\left(1+\frac{i
      \tau_k}{8}\right)+\sqrt{\frac{\tau_k}{8}}\Gamma\left(\frac{1}{2}+\frac{i
      \tau_k}{8}\right) e^{i \pi/4}\right|^2
\eeq
with
\be
\tau_k=\frac{4  \Delta_k^2}{ \delta_V}.
\ee
To find the total density of excited particles we have to sum the probability $p_{\rm ex}(k)$ over all the momenta in the first Brillouin zone:
\be
n_{\rm ex}={1\over L}\sum_{k\in[-\pi/a,\pi/a]} p_{\rm ex}(k).
\ee
 It is convenient to introduce a shifted momentum $q=\left(k-k_f\right)$. Then we have  $\tau_k=(q/k_\delta)^2$ with $k_\delta=\frac{\sqrt{\delta_V}}{2}$. In the limit of small $\delta_V$ the upper limit of the integral over $q$ can be sent to infinity, therefore
\be
n_{\rm ex}\approx  2 \int_{0}^\infty \frac{dq}{2\pi}p_{\rm ex}(q^2/k_\delta^2)\approx 0.12 k_\delta.
\ee
We see that the exact solution confirms the general scaling $n_{\rm ex} \propto \sqrt{\delta}$ (see Eq.~(\ref{n_ex1})) provided that $d=z=\nu=1$.

This problem can be also solved perturbatively. In fact the loading problem becomes equivalent to the transverse field Ising model. For example from the mapping to the LZ problem it easy to see that the matrix element appearing in   Eq.~(\ref{int_eq3}) is:
\be
\langle +_k|\partial_\lambda|-_k\rangle=\frac{1}{2}\frac{2 \Delta_k}{(2 \Delta_k)^2+V^2}
\ee
which is similar to Eq.~(\ref{el_ising}). The situation is also analogous for the energy spectrum. The net result of the calculation is
\be
n^{\rm pert}_{\rm ex}  \approx  0.14 k_\delta.
\ee
The results for the entropy and heat for the loading problem have identical scaling with the loading rate.

\section{Sudden quenches near quantum critical points}

So far we focused exclusively on the adiabatic regime of small $\delta$. For simplicity we restrict the discussion in this section only to the situation where we start at the critical point and increase $\lambda$ in time: $\lambda(t)=\delta t$. We argued that in low dimensions, when $d\nu/(z\nu+1)<2$, the dynamics is dominated by the low energy excitations with energies $\epsilon\lesssim \epsilon^\star=|\delta|^{z\nu/(z\nu+1)}$. The energy scale $\epsilon^\star$ corresponds to the value of the tuning parameter $\lambda^\star=|\delta|^{1/(z\nu+1)}$ (because of the general scaling relation $\epsilon\sim \lambda^{z\nu}$). So equivalently we can say that in low dimensions the non-adiabatic effects are dominated by the transitions occurring in the vicinity of the critical point $|\lambda|\lesssim \lambda^\star$.  A simple qualitative way to understand the adiabatic dynamics in this case is to split the time evolution into two domains: $t<t^\star=\lambda^\star/\delta\sim 1/|\delta|^{z\nu/(z\nu+1)}$ and $t>t^\star$. Then in the first domain the dynamics can be thought as approximately fast (sudden) and in the second domain the dynamics is adiabatic (see also discussion in Ref.~\cite{claudia}). Then we can think about the slow quench as a sudden quench with the amplitude of the quench being equal to $\lambda^\star$. I.e. one can expect that the scaling of the density of excitations and other thermodynamic quantities can be approximately obtained by projecting the initial ground state corresponding to the critical point $(\lambda=0)$ to the eigenstates of the new quenched Hamiltonian with $\lambda=\lambda^\star$. Such approach was indeed successfully applied to the problem of quenching the system through the BCS-BEC crossover~\cite{ehud_ashvin}. This argument implies that for sudden quenches we should have that $n_{\rm ex}\sim |\lambda^\star|^{d\nu}$~\cite{claudia_quench}. For $d\nu<2$ this is a non-analytic function of the quench rate, which can not be obtained within the ordinary perturbation theory. As we will show below this scaling, however, immediately follows from the adiabatic perturbation theory applied to sudden quenches. Also we will show  that when $d\nu>2$ this scaling fails and the perturbative analytic result is restored $n_{\rm ex}\sim (\lambda^\star)^2$. As in the case of slow quenches the analytic quadratic dependence comes from the dominance of the high energy excitations~\cite{claudia_quench}.

Let us return to the discussion of Sec.~\ref{Sec:pert}, where we introduced the adiabatic perturbation theory. We emphasize that the word adiabatic only means that we are working in the instantaneous (adiabatic) basis. The small parameter in this theory is the probability to excite higher energy levels. For slow quenches the excitation probability is small because the rate $\delta$ is small, while for fast quenches this probability is small because the quench amplitude $\lambda^\star$ is small. Thus for sudden quenches of small amplitude we can still use Eq.~(\ref{int_eq3}), with further simplification that the phase factor $\Theta_n-\Theta_0\to 0$ since it is inversely proportional to the rate $\delta\to\infty$. Thus instead of Eq.~(\ref{int_eq3}) we can write~\cite{claudia_quench}
 \be
 \alpha_n(\lambda^\star)\approx-\int\limits_{0}^{\lambda^\star} d\lambda' \langle
 n|\partial_{\lambda'}|0\rangle=\int\limits_{0}^{\lambda^\star} d\lambda' {\langle
 n|V|0\rangle\over E_n(\lambda')-E_0(\lambda')}.
 \label{int_eq4}
 \ee
Note that in the case where there is a finite gap in the spectrum, the difference $E_n(\lambda')-E_0(\lambda')$ remains large for all values of $\lambda'\in[0,\lambda^\star]$. If the same is true for the matrix element then this expression reduces to the one from the ordinary perturbation theory:
\be
\alpha_n(\lambda^\star)\approx \lambda^\star {\langle
 n|V|0\rangle_{\lambda=0}\over E_n(0)-E_0(0)}.
 \label{lr}
\ee
The advantage of using the expression~(\ref{int_eq4}) for $\alpha_n$ over the standard perturbative result (\ref{lr}) is that Eq.~(\ref{int_eq4}) does not give an explicit preference to the initial state $\lambda=0$ over the final state $\lambda=\lambda^\star$ and it does not assume the analytic behavior of $\alpha_n$ with $\lambda^\star$. As we will see both these points are necessary for studying quench dynamics near QCP.

Let us apply Eq.~(\ref{int_eq4}) to the problem of quenching starting  from the critical point. Using the scaling ansatz for the matrix element (\ref{scaling_matr_el}), we find  that the density of excitations is approximately given by
\be\label{eq:n_ex_scaling}
n_{\rm ex}\approx \int \frac{d^d\,k}{(2\pi)^d} \left| \int_0^{\lambda^\star} \frac{d\lambda}{\lambda} \, G\left(\frac{k}{\lambda^{\nu}}\right)\right|^2.
\ee
To analyze this expression it is convenient to change variables as: $\lambda=\lambda^\star\xi$, $k=(\lambda^\star)^\nu\eta$. Then we obtain
\be\label{n_sudden2}
n_{\rm ex}\approx |\lambda^\star|^{d\nu} \int \frac{d^d\,\eta}{(2\pi)^d} \left| \int_0^{1} \frac{d\xi}{\xi} \, G\left(\frac{\eta}{\xi^{\nu}}\right)\right|^2.
\ee
This expression gives the desired scaling $n_{\rm ex}\sim (|\lambda|^\star)^{d\nu}$ provided that the integral over $\eta$ converges at large $\eta$. From the asymptotics of the scaling function $G(x)\propto 1/x^{1/\nu}$ at large $x$, discussed earlier, we see that the necessary condition for convergence of the integral is $d<2/\nu$ or $d\nu<2$. In the opposite limit, $d\nu>2$,  the high energy quasiparticles dominate the total energy and one can use the ordinary perturbation theory (linear response) which gives
\be
n_{ex}^{lr}\approx |\lambda^\star|^2 \int {d^dk\over (2\pi)^d}{\left|\langle k| V|0\rangle_0\right|^2\over |\varepsilon_k^0-\varepsilon_0^0|^2},
\ee
where all the quantities are evaluated at the critical point $\lambda=0$. We note that the quantity multiplying $|\lambda^\star|^2$ is called the fidelity susceptibility~\cite{gu}. Thus the regime of validity of the universal scaling (\ref{n_sudden2}) corresponds to  a divergent fidelity susceptibility. One can make similar analysis of the scaling of the heat and the entropy:
\be
Q\propto |\lambda^\star|^{(d+z)\nu},\quad S_d\propto |\lambda^\star|^{d\nu}.
\ee
As in the case of $n_{\rm ex}$ these scaling laws are valid as long as the corresponding exponents are less than two. One can verify that these scalings are indeed reproduced for the models we analyzed in the previous section. Both adiabatic perturbation theory and the exact calculation give the same scaling, however, the perturbative approach gives a mistake in the prefactor.

From the above analysis we see that in low dimensions $d\nu<2$ there is indeed a direct analogy between slow and sudden quenches. One gets the same scaling if correctly identifies the quench rate $\delta$ in the former case and the quench amplitude $\lambda^\star$ in the latter: $\lambda^\star\sim |\delta|^{1/(z\nu+1)}$. This analogy has direct similarity with the original arguments by Kibble and Zurek who predicted the scaling (\ref{n_ex1}) in the context of topological defect formation while crossing classical phase transition with the parameter $\delta$ playing the role of temperature quenching rate~\cite{kz1, kz2}. The main argument in the Kibble-Zurek (KZ) mechanism is that there is a divergent relaxation time scale near the critical point and a corresponding divergent length scale. The topological excitations with distances bigger than this length scale do not have a chance to thermalize and remain in the system for a long time even if they are thermodynamically forbidden. A simple estimate gives that this length scale behaves precisely as $\xi\sim 1/|\delta|^{\nu/(z\nu+1)}$~\cite{kz2}, resulting in a density of topological defects $n_{\rm ex}\sim 1/\xi^d$, which is the same as in Eq.~(\ref{n_ex1}). One can thus think about the KZ mechanism as adiabatic quenching of the temperature up to the scale $T^\star$, corresponding to the correlation length $\xi$, followed by sudden quench of the temperature to zero, so that remaining topological excitations essentially freeze. Essentially the same argument we used here in the quantum case and it indeed works qualitatively right for $d\nu<2$.

The situation is different in the regime $d\nu/(z\nu+1)<2<d\nu$. In this case for  slow quenches the universal scaling is still applicable, i.e. only the low energy excitations created at $\lambda\lesssim \lambda^\star$ are important for the scaling of $n_{\rm ex}$ while for sudden quenches this is no longer the case. Namely, the scaling of $n_{\rm ex}$ with the quench amplitude becomes quadratic and non-universal, i.e. sensitive to the high energy cutoff. Therefore in this regime the analogy between slow and sudden quenches becomes misleading. This makes an important difference with the Kibble-Zurek mechanism where such issues do not arise. If $d\nu/(z\nu+1)>2$ then the scaling becomes quadratic in both cases (with the rate $\delta$ for slow quenches and with the amplitude $\lambda^\star$ for sudden quenches), with the main contribution to excitations coming from the high energy quasiparticles.

\section{Effect of the quasiparticle statistics in the finite temperature quenches}

One can try to extend the analysis we carried through (both for sudden and slow quenches) to the finite temperature situation. We note that because we consider isolated systems with no external bath, the temperature only describes the initial density matrix. To realize this situation one can imagine that the system weakly couples to a thermal bath and reaches some thermal equilibrium. Then, on the time scales of the dynamical processes we are interested in, this coupling has a negligible effect and the dynamics of the system is essentially Hamiltonian. One does not even have to assume a coupling to the external bath if we are dealing with ergodic systems, since they are believed to reach thermal equilibrium states by themselves. By now this problem remains unsolved in the most general case. As we will argue below at finite temperatures the statistics of the low energy quasiparticles plays a key role. In general one can have critical points where the low energy excitations have fractional statistics or do not have well defined statistics at all.

In this section we will consider only a relatively simple situation in which the statistics of quasiparticles is either bosonic or fermionic. In particular, the two limits of the sine-Gordon model corresponding to $K_{SG}\ll 1$, where the system maps to a set of independent harmonic oscillators, and $K_{SG}=1$, where the system is equivalent to non-interacting fermions, are examples of critical systems with bosonic and fermionic statistics respectively. The other two examples in this paper (harmonic chain and transverse field Ising model) obviously also fall into the category of a system with well defined statistics of quasiparticles (bosonic and fermionic respectively).

For bosonic excitations it is straightforward to show that the Gaussian ansatz (\ref{gauss}) still holds at finite temperatures for the Wigner function~\cite{pg_np, ap_wig}. The width of the Wigner function satisfies a similar equation as the width of the wave function~(\ref{eq:sig}), (\ref{eq:sig1}). The only difference with the initial ground state is that this width gets ``dressed'' as  $\sigma_k\coth(\epsilon^0_k/2T)$ at initial time, where $\sigma_k^0$ is the ground state width defined earlier and $\epsilon_k^0$ is the initial energy of the particle with  momentum $k$. Using this fact it is straightforward to show~\cite{pg_np} that:
\bed
\frac{1}{2} \left[\frac{\sigma_k^{\mathrm{eff}}}{\sigma_k^{\mathrm{eq}}} -1\right]\longrightarrow\frac{1}{2} \left[\frac{\sigma_k^{\mathrm{eff}}}{\sigma_k^{\mathrm{eq}}}\coth\left(\frac{\epsilon^0_k}{2T}\right) -1\right].
\eed
Since we are interested in the excitations created by the dynamical process, we need to subtract from this quantity the initial number of quasiparticles excitations, which were present due to initial thermal fluctuations:
\beq
&&n^T_{\rm ex}(k) = \frac{1}{2} \left[\frac{\sigma_k^{\mathrm{eff}}}{\sigma_k^{\mathrm{eq}}}\coth\left(\frac{\epsilon^0_k}{2T}\right) -1\right]\nonumber\\
&&-\frac{1}{2} \left[\coth\left(\frac{\epsilon^0_k}{2T}\right) -1\right]= n^0_{\rm ex}(k) \coth\left(\frac{\epsilon_k^0}{2T}\right),\phantom{XX}
\label{nex_bos_T}
\eeq
where $n^0_{\rm ex}(k)$ is the number of created quasiparticles with momentum $k$ at zero temperature. Note that this result does not depend on the details of the quench process, i.e. whether it is fast or slow. At low temperatures $T\ll \epsilon_k^0$ the expression (\ref{nex_bos_T}) obviously reduces to the zero temperature result. However, at high temperatures $T\gg \epsilon_k^0$ we have the bosonic enhancement of the transitions
\be
n^T_{\rm ex}(k)\approx n^0_{\rm ex}(k) {2T\over \epsilon_k^0}.
\ee
When we sum $n^T_{\rm ex}(k)$ over all possible momenta $k$ starting from a gapless system, this extra $T/\epsilon_k^0$ factor makes the result more infrared divergent. In the regime of validity of the adiabatic perturbation theory it changes the scaling for example for the density of excitations to
\be
n_{\rm ex}\propto T|\delta|^{(d-z)\nu/(z\nu+1)}.
\label{sc_bos_T}
\ee
The crossover, where the quadratic scaling is restored, is determined
by the  equation $(d-z)\nu/(z\nu+1)=2$. We note, however, that in small dimensions the response of the system can become non-adiabatic~\cite{pg_np} and the scaling above can break down. For the sine-Gordon model in the bosonic limit using Eq.~(\ref{nk2}) it is indeed straightforward to see that
\be
n_{\rm ex}\sim T(c\delta)^{1/3}L,
\ee
where $L$ is the system size. If we consider massive bosonic theories in higher dimensions we would find that
the scaling (\ref{sc_bos_T}) would be restored above two dimensions. Even though in one dimension the adiabatic perturbation theory fails to predict the correct scaling, it unambiguously shows that the  finite temperatures make the response of the system less adiabatic. For sudden quenches the effect of initial temperature on harmonic system was recently considered in Ref.~\cite{cardy_temp} and the authors obtained results consistent with the statements above.

The scenario becomes quite opposite in the fermionic case, $K_{SG}=1$, where we are dealing with a sum of independent two-level systems. At finite temperature each level is occupied according to the Fermi distribution
\begin{displaymath}
f_k^{\pm }=\left(\exp\left[\pm \frac{ \epsilon_k^0}{T}\right]+1\right)^{-1}.
\end{displaymath}
The probability of excitation thus gets corrected as:
\be
n^T_{\rm ex}(k) = n^0_{\rm ex}(k) (f_k^{-}-f_k^{+}) = n^0_{\rm
ex}(k)\tanh\left(\frac{ \epsilon_k^0}{2 T}\right).
\ee
The fact that we got $tanh$  factor for fermions (and $coth$ for bosons) is hardly surprising. Similar factors appear in conventional fluctuation-dissipation relations~\cite{abrikosov}. However, this mere change of $coth$ to $tanh$ factor has an important implication. As in the case of bosons, in the small temperature limit this additional factor reduces to unity and the zero temperature result is recovered. At high temperatures $T\gg \epsilon_k^0$ we find
\be
n^T_{\rm ex}(k)\approx n^0_{\rm ex}(k) \frac{ \epsilon_k^0}{2 T}.
\ee
Therefore the number of created quasiparticles is much smaller than in the zero temperature case. This fact reflects fermionic anti-bunching. In other words the preexisting thermal quasiparticles are blocking the transition to the already occupied excited states and the dynamical process becomes more adiabatic. Thus for the density of excitations we expect now the scaling
\be
n_{\rm ex}\propto |\delta|^{(d+z)\nu/(z\nu+1)}/T.
\label{sc_ferm_T}
\ee

This analysis is not sensitive to the details of the process, hence the situation remains the same for sudden quenches. Thus instead of the scaling $n_{\rm ex}\sim |\lambda^\star|^{d\nu}$ we will get $n_{\rm ex}\sim |\lambda^\star|^{(d\mp z)\nu} T^{\pm 1}$ where the upper ``+'' or ``--'' sign corresponds to bosons and the lower sign does to fermions. The crossover to the quadratic linear response scaling happens when the corresponding exponent becomes two. As in the case of slow quenches one should be careful with the validity of the perturbative scaling in low dimensions for bosonic excitations, where the system size can affect the scaling and change the exponent.

The main conclusion of this section is that at finite initial temperatures the statistic of quasiparticles qualitatively changes the response of the quantum critical system to quenches (independently if  they are slow or fast). At zero temperature the quasiparticle statistics does not seem to play an important role, since it does not enter the general scaling exponents. This suggests that the non-adiabatic response will be extremely interesting in the systems with fractional statistics of excitations. This sensitivity and potential universality of the response of the system to sudden or slow perturbations at finite temperatures might allow one to use non-adiabatic transitions as an experimental probe of the quasiparticle statistics.

\section{Conclusions}

In this work we focused on the analysis of the response of a translationally invariant system, initially prepared in the ground state, to linear quenches, where an external parameter globally coupled to the whole system linearly changes in time. Using the adiabatic perturbation theory we showed how to obtain the scaling of various quantities like the density of quasiparticles $n_{\rm ex}$, heating (excess non-adiabatic energy) $Q$ and entropy $S_d$  with the quench rate $\delta$, for small $\delta$. We started from a simple two-level system, where we showed that the transition probability scales quadratically with $\delta$, at $\delta\to 0$, if the external parameter changes in the finite range. We then showed that this quadratic scaling can be violated in various low-dimensional systems, especially quenched through singularities like quantum critical points. This violation comes because excitations of low energy levels, for which dynamics is diabatic, dominate the scaling of $n_{\rm ex}$ and other quantities, which then acquire universal non-analytic dependence on the rate $\delta$. For example, we argued that in generic low-dimensional gapless systems $n_{\rm ex}\propto |\delta|^{d/z}$ as long as $d/z<2$. In the opposite limit $d/z>2$ we expect that the high energy quasiparticles dominate $n_{\rm ex}$ and the quadratic scaling is restored: $n_{\rm ex}\propto \delta^2$. Similar story is true for heat and entropy. We note that the quadratic scaling of the transition probability to highly excited states is specific to linear quenches, where the time derivative of the tuning parameter has a discontinuity at initial and final times. E.g. we analyzed the situation where $\dot\lambda=\delta$ for $t>t_i$ and $\dot\lambda=0$ for $t<t_i$. If the tuning parameter is turned on and off smoothly in time then this quadratic scaling is no longer valid. One can argue, however, that at sufficiently high energies there are always dephasing mechanisms for quasiparticles which reset their phase and effectively reset the value of $t_i$. This resetting of the phase will likely restore the quadratic scaling. However, this issue needs to be investigated separately and is beyond the scope of this work.

Using the same adiabatic perturbation theory and the general scaling arguments we showed how the universal behavior of $n_{\rm ex}$, $S_d$, and $Q$ emerges for quenching through quantum critical points. In particular, we showed that the density of generated quasiparticles (and entropy) scale as $n_{\rm ex}, S_d\sim |\delta|^{d\nu/(z\nu+1)}$, where $\nu$ is the critical exponent characterizing the divergence of the correlation length near the phase transition in agreement with earlier works~\cite{ap_adiabatic, zurek_adiabatic}. Based on the scaling analysis we showed that for linear quenches this scaling is valid as long as $d\nu/(z\nu+1)<2$, otherwise the quadratic scaling is restored. We also discussed the connection between adiabatic and sudden quenches near a quantum critical point. We argued that if $d\nu<2$ the two are qualitatively similar provided that one correctly associates the quench amplitude $\lambda^\star$ and the quench rate $\delta$: $\lambda^\star\sim |\delta|^{1/(z\nu+1)}$. Using the adiabatic perturbation theory adopted to sudden quenches of small amplitude we showed that the density of excitations scales with the quench amplitude as $n_{\rm ex}\propto |\lambda^\star|^{d\nu}$. When $d\nu>2$  the quadratic scaling, which follows from the conventional perturbation theory (linear response), is restored and $n_{\rm ex}\sim (\lambda^\star)^2$. As in the case of slow quenches the quadratic scaling comes from the dominant contribution of the high energy quasiparticles to $n_{\rm ex}$.

We also discussed the situation where a gapless system is initially prepared at finite temperature. In this case we argued that the statistics of the quasiparticles becomes crucial. In particular, for bosonic quasiparticles the finite temperature enhances the effects of non-adiabaticity due to bunching effect, while conversely for fermions the response becomes more adiabatic than at zero temperature due to the Pauli blocking. In the regime of validity of the adiabatic perturbation theory we argued that in the scaling laws presented above (both for sudden and slow quenches) one should change $d\to d-z$ in the bosonic case and $d\to d+z$ in the fermionic case. We note, however, that in the bosonic case the adiabatic perturbation theory can breakdown due to overpopulation of low energy bosonic modes and the system can enter a non-adiabatic regime~\cite{pg_np}. We illustrated  our general statements with explicit results for various solvable models.

{\em Acknowledgements.} We would like to acknowledge R.~Barankov and V.~Gritsev for useful discussions and comments. We also would like to acknowledge R.~Barankov for providing Fig.~\ref{fig:Higgs_diagram} and help in derivation of Eq.~(\ref{eq:LZ_int}).  This work was supported by AFOSR YIP and Sloan Foundation. C.~D.~G. acknowledges the support of I2CAM: DMR-0645461.

\appendix

\section{\emph{half} Landau-Zener problem}\label{LZ_app}

\begin{center}
\begin{figure}[ht]
\includegraphics[width=3.2in]{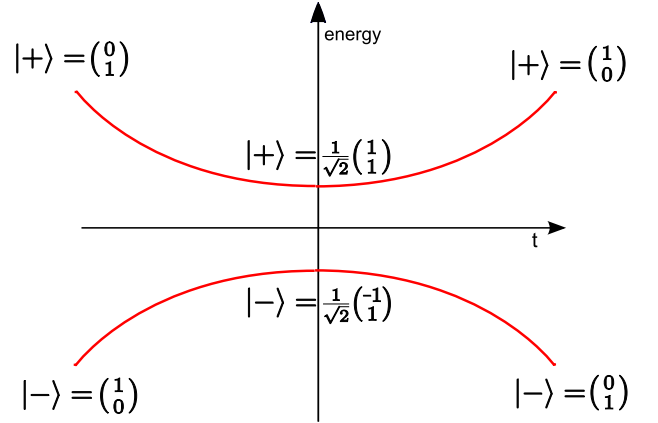}
\caption{Landau-Zener problem: the eigenvectors of the system are exchanged during the time evolution.}
\label{landau_zener:fig}
\end{figure}
\end{center}

Here we briefly describe the derivation of the transition probability in the LZ problem, when the system starts at the symmetric point with the smallest gap. Let us consider the Hamiltonian:
\be \label{landau_zener:h}
H=\left[\begin{matrix} \lambda(t) & g \\ g & -\lambda(t)\end{matrix}\right],
\ee
where $g$ is a  constant and  $\lambda(t)=\delta t$. This Hamiltonian has normalized eigenvectors given by Eq.~(\ref{eigen_lz}):
\be
|-\rangle=\left(\begin{array}{c} \sin(\theta/2) \\ -\cos(\theta/2)\end{array}
\right),\quad |+\rangle=\left(\begin{array}{c} \cos(\theta/2) \\
\sin(\theta/2)\end{array}\right),
\ee
where $\tan\theta(t)=g/\lambda(t)$, corresponding to the eigenergies
\be
E_{\pm}(t)=\pm \sqrt{(\lambda(t))^2+ g^2}.
\ee
The vectors representing the ground state $\vert - \rangle$ and the excited state $\vert + \rangle$ significantly
change as the time $t$ is varied, as it is sketched in Fig.\ref{landau_zener:fig}.

Unlike in Sec.~\ref{sec:lz} for the exact solution it is convenient to work in the fixed basis:
\be
\vert \psi(t) \rangle =\phi_1(t)\left(\begin{array}{c} 1 \\ 0 \end{array} \right)+ \phi_2(t) \left(\begin{array}{c} 0 \\ 1 \end{array} \right). \no
\ee
Then the equations of motion for the coefficients $\phi_1(t)$ and $\phi_2(t)$ become
\begin{eqnarray}\label{eq_phi1}
i \dot{\phi_1}=\delta t \phi_1 + g \phi_2  \\
\label{eq_phi2}
i  \dot{\phi_2}=g \phi_1 -\delta t \phi_2.
\end{eqnarray}
This system of equations has the following generic solutions for $\phi_1$ and  $\phi_2$:
\begin{widetext}
\be\label{LZsolut1}
\phi_1(t)=c_1 e^{-i\frac{t^2}{4}}
M\left(\frac{i g^2}{4 \delta},\frac{1}{2},\frac{i
    t^2}{2}\right)+c_2 t  e^{-i\frac{t^2}{4}}
M\left(\frac{1}{2}+\frac{i g^2}{4 \delta},\frac{3}{2},\frac{i t^2}{2}\right),
\ee
\be\label{LZsolut2}
\phi_2(t)=c_3 e^{-i\frac{t^2}{4}}
M\left(\frac{1}{2}+\frac{i g^2}{4 \delta},\frac{1}{2},\frac{i
    t^2}{2}\right)+c_4 t  e^{-i\frac{t^2}{4}}
M\left(1+\frac{i g^2}{4 \delta},\frac{3}{2},\frac{i t^2}{2}\right),
\ee
\end{widetext}
where $M(a,b,z)$ is the confluent hypergeometric function otherwise called
$_1 F_1$. The coefficients $c_1,c_2,c_3,c_4$ are determined by the initial conditions on the wave function
at the initial time $t_{\rm in}$
\be
\vert \psi(t_{\rm in}) \rangle =\phi_1(t_{\rm in})\left(\begin{array}{c} 1 \\ 0 \end{array} \right)+ \phi_2(t_{\rm in}) \left(\begin{array}{c} 0 \\ 1 \end{array} \right) \no
\ee
plus two auxiliary conditions e.g.
\begin{eqnarray}\label{cond0}
i \dot{\phi_1}\rvert_{t=0}=\frac{g}{ \sqrt{2 \delta}} \phi_2\rvert_{t=0} \\
\label{cond1}
i  \dot{\phi_2}\rvert_{t=0}=\frac{g}{ \sqrt{2 \delta}}\phi_1\rvert_{t=0}.
\end{eqnarray}
The conventional LZ problem ~\cite{lz1, lz2} considers the case of starting in the ground state  at $t_{\rm}=-\infty$ and asks what is the excitation probability after evolving the system to a final time $t=+\infty$, which is
\be
p_{\rm ex}=\exp\left[-{\pi g^2\over \delta}\right].
\ee
This result can be recovered imposing the initial conditions:
\be
t_{\rm in}=-\infty \quad \vert \psi(- \infty) \rangle =\vert - \rangle=\left(\begin{array}{c} 1 \\ 0 \end{array} \right)\quad \begin{array}{c} \phi_1(-\infty)=1 \\ \phi_2(-\infty)=0 \end{array}\no
\ee
and looking at the asymptotic behavior of $\vert \phi_1(\infty) \vert^2$.

Another situation that can be straightforwardly considered corresponds to starting from $t=0$ (instead of $t=-\infty$) and ending at $t=+\infty$, essentially this is \emph{half} of the usual LZ-problem. As we have explained in Sec.~\ref{sec:sg} this situation naturally arises in e.g. describing the problem of loading hard-core bosons into an optical lattice. Here we need to impose the initial conditions
\be
t_{\rm in}=0 \quad \vert \psi(0) \rangle =\vert - \rangle=\frac{1}{\sqrt{2}}\left(\begin{array}{c} -1 \\ 1 \end{array} \right)\quad \begin{array}{c} \phi_1(0)=-\frac{1}{\sqrt{2}} \\ \phi_2(0)=\frac{1}{\sqrt{2}} \end{array}.\no
\ee
The probability of excitation $p_{\rm ex}=\lim_{t \to \infty} \vert \phi_1(\infty) \vert^2$
turns out to be (see also Refs.~\cite{vitanov_99, damski_zurek})
\beq \label{probtauLZ}
p_{\rm ex}(\tau)& = & 1-2\frac{ e^{-\pi \tau/8}}{\pi
  \tau}\sinh(\frac{\pi \tau}{4})  \\
 & &  \times \left|\Gamma\left(1+\frac{i
      \tau}{8}\right)+\sqrt{\frac{\tau}{8}}\Gamma\left(\frac{1}{2}+\frac{i
      \tau}{8}\right) e^{i \pi/4}\right|^2.\no
\eeq
where we introduced  $\tau= 2 g^2/\delta$. This function has the following asymptotic behavior: for  $\tau \to 0$
 \begin{displaymath}
 p_{\rm ex}\approx \frac{1}{2}-\frac{\sqrt{\pi \tau}}{4}
 \end{displaymath}
 and for $\tau \to \infty$
 \begin{displaymath}
 p_{\rm ex} \approx \frac{1}{4 \tau^2}=\frac{\delta^2}{16 g^4}.
 \end{displaymath}
 As expected the latter slow asymptotic agrees with the result of the adiabatic perturbation theory (\ref{tr_prob_lz_sym}).

\bibliography{kolkata}

\begin{thebibliography}{70}
\expandafter\ifx\csname natexlab\endcsname\relax\def\natexlab#1{#1}\fi
\expandafter\ifx\csname bibnamefont\endcsname\relax
  \def\bibnamefont#1{#1}\fi
\expandafter\ifx\csname bibfnamefont\endcsname\relax
  \def\bibfnamefont#1{#1}\fi
\expandafter\ifx\csname citenamefont\endcsname\relax
  \def\citenamefont#1{#1}\fi
\expandafter\ifx\csname url\endcsname\relax
  \def\url#1{\texttt{#1}}\fi
\expandafter\ifx\csname urlprefix\endcsname\relax\def\urlprefix{URL }\fi
\providecommand{\bibinfo}[2]{#2}
\providecommand{\eprint}[2][]{\url{#2}}

\bibitem[{\citenamefont{Bloch et~al.}(2008)\citenamefont{Bloch, Dalibard, and
  Zwerger}}]{Bloch2008_rmp}
\bibinfo{author}{\bibfnamefont{I.}~\bibnamefont{Bloch}},
  \bibinfo{author}{\bibfnamefont{J.}~\bibnamefont{Dalibard}}, \bibnamefont{and}
  \bibinfo{author}{\bibfnamefont{W.}~\bibnamefont{Zwerger}},
  \bibinfo{journal}{Reviews of Modern Physics} \textbf{\bibinfo{volume}{80}},
  \bibinfo{pages}{885} (\bibinfo{year}{2008}).

\bibitem[{\citenamefont{Greiner et~al.}(2002)\citenamefont{Greiner, Mandel,
  Hansch, and Bloch}}]{Greiner2002b}
\bibinfo{author}{\bibfnamefont{M.}~\bibnamefont{Greiner}},
  \bibinfo{author}{\bibfnamefont{O.}~\bibnamefont{Mandel}},
  \bibinfo{author}{\bibfnamefont{T.~W.} \bibnamefont{Hansch}},
  \bibnamefont{and} \bibinfo{author}{\bibfnamefont{I.}~\bibnamefont{Bloch}},
  \bibinfo{journal}{Nature} \textbf{\bibinfo{volume}{419}}, \bibinfo{pages}{51}
  (\bibinfo{year}{2002}).

\bibitem[{\citenamefont{Fertig et~al.}(2005)}]{fertig2005}
\bibinfo{author}{\bibfnamefont{C.~D.} \bibnamefont{Fertig}}
  \bibnamefont{et~al.}, \bibinfo{journal}{Phys. Rev. Lett.}
  \textbf{\bibinfo{volume}{94}}, \bibinfo{pages}{120403}
  (\bibinfo{year}{2005}).

\bibitem[{\citenamefont{Tuchman et~al.}(2006)\citenamefont{Tuchman, Orzel,
  Polkovnikov, and Kasevich}}]{ari}
\bibinfo{author}{\bibfnamefont{A.~K.} \bibnamefont{Tuchman}},
  \bibinfo{author}{\bibfnamefont{C.}~\bibnamefont{Orzel}},
  \bibinfo{author}{\bibfnamefont{A.}~\bibnamefont{Polkovnikov}},
  \bibnamefont{and} \bibinfo{author}{\bibfnamefont{M.}~\bibnamefont{Kasevich}},
  \bibinfo{journal}{Phys. Rev. A} \textbf{\bibinfo{volume}{74}},
  \bibinfo{pages}{051601} (\bibinfo{year}{2006}).

\bibitem[{\citenamefont{Kinoshita et~al.}(2006)\citenamefont{Kinoshita, Wenger,
  and Weiss}}]{newtoncradle}
\bibinfo{author}{\bibfnamefont{T.}~\bibnamefont{Kinoshita}},
  \bibinfo{author}{\bibfnamefont{T.}~\bibnamefont{Wenger}}, \bibnamefont{and}
  \bibinfo{author}{\bibfnamefont{D.~S.} \bibnamefont{Weiss}},
  \bibinfo{journal}{Nature} \textbf{\bibinfo{volume}{440}},
  \bibinfo{pages}{900} (\bibinfo{year}{2006}).

\bibitem[{\citenamefont{Sadler et~al.}(2006)\citenamefont{Sadler, Higbie,
  Leslie, Vengalattore, and Stamper-Kurn}}]{sadler}
\bibinfo{author}{\bibfnamefont{L.~E.} \bibnamefont{Sadler}},
  \bibinfo{author}{\bibfnamefont{J.~M.} \bibnamefont{Higbie}},
  \bibinfo{author}{\bibfnamefont{S.~R.} \bibnamefont{Leslie}},
  \bibinfo{author}{\bibfnamefont{M.}~\bibnamefont{Vengalattore}},
  \bibnamefont{and} \bibinfo{author}{\bibfnamefont{D.~M.}
  \bibnamefont{Stamper-Kurn}}, \bibinfo{journal}{Nature}
  \textbf{\bibinfo{volume}{443}}, \bibinfo{pages}{312} (\bibinfo{year}{2006}).

\bibitem[{\citenamefont{Hofferberth et~al.}(2007)\citenamefont{Hofferberth,
  Lesanovsky, Fischer, Schumm, and Schmiedmayer}}]{schumm2007}
\bibinfo{author}{\bibfnamefont{S.}~\bibnamefont{Hofferberth}},
  \bibinfo{author}{\bibfnamefont{I.}~\bibnamefont{Lesanovsky}},
  \bibinfo{author}{\bibfnamefont{B.}~\bibnamefont{Fischer}},
  \bibinfo{author}{\bibfnamefont{T.}~\bibnamefont{Schumm}}, \bibnamefont{and}
  \bibinfo{author}{\bibfnamefont{J.}~\bibnamefont{Schmiedmayer}},
  \bibinfo{journal}{Nature} \textbf{\bibinfo{volume}{449}},
  \bibinfo{pages}{324} (\bibinfo{year}{2007}).

\bibitem[{\citenamefont{Trotzky et~al.}(2008)}]{Trotzky2008}
\bibinfo{author}{\bibfnamefont{S.}~\bibnamefont{Trotzky}} \bibnamefont{et~al.},
  \bibinfo{journal}{Science} \textbf{\bibinfo{volume}{319}},
  \bibinfo{pages}{295} (\bibinfo{year}{2008}).

\bibitem[{\citenamefont{Rigol et~al.}(2008)\citenamefont{Rigol, Dunjko, and
  Olshanii}}]{olshanii_nature}
\bibinfo{author}{\bibfnamefont{M.}~\bibnamefont{Rigol}},
  \bibinfo{author}{\bibfnamefont{V.}~\bibnamefont{Dunjko}}, \bibnamefont{and}
  \bibinfo{author}{\bibfnamefont{M.}~\bibnamefont{Olshanii}},
  \bibinfo{journal}{Nature} \textbf{\bibinfo{volume}{452}},
  \bibinfo{pages}{854} (\bibinfo{year}{2008}).

\bibitem[{\citenamefont{Reimann}(2008)}]{reimann}
\bibinfo{author}{\bibfnamefont{P.}~\bibnamefont{Reimann}},
  \bibinfo{journal}{Phys. Rev. Lett.} \textbf{\bibinfo{volume}{101}},
  \bibinfo{pages}{190403} (\bibinfo{year}{2008}).

\bibitem[{\citenamefont{Silva}(2008)}]{silva}
\bibinfo{author}{\bibfnamefont{A.}~\bibnamefont{Silva}},
  \bibinfo{journal}{Phys. Rev. Lett.} \textbf{\bibinfo{volume}{101}},
  \bibinfo{pages}{236803} (\bibinfo{year}{2008}).

\bibitem[{\citenamefont{Rigol}(unpublished)}]{rigol2009_therm}
\bibinfo{author}{\bibfnamefont{M.}~\bibnamefont{Rigol}},
  \bibinfo{journal}{arXiv:0904.3746}  (\bibinfo{year}{unpublished}).

\bibitem[{\citenamefont{Altman and Auerbach}(2002)}]{ehud_assa}
\bibinfo{author}{\bibfnamefont{E.}~\bibnamefont{Altman}} \bibnamefont{and}
  \bibinfo{author}{\bibfnamefont{A.}~\bibnamefont{Auerbach}},
  \bibinfo{journal}{Phys. Rev. Lett.} \textbf{\bibinfo{volume}{89}},
  \bibinfo{pages}{250404} (\bibinfo{year}{2002}).

\bibitem[{\citenamefont{Polkovnikov et~al.}(2002)\citenamefont{Polkovnikov,
  Sachdev, and Girvin}}]{psg}
\bibinfo{author}{\bibfnamefont{A.}~\bibnamefont{Polkovnikov}},
  \bibinfo{author}{\bibfnamefont{S.}~\bibnamefont{Sachdev}}, \bibnamefont{and}
  \bibinfo{author}{\bibfnamefont{S.~M.} \bibnamefont{Girvin}},
  \bibinfo{journal}{Phys. Rev. A} \textbf{\bibinfo{volume}{66}},
  \bibinfo{pages}{053607} (\bibinfo{year}{2002}).

\bibitem[{\citenamefont{Barankov et~al.}(2004)\citenamefont{Barankov, Levitov,
  and Spivak}}]{barankov}
\bibinfo{author}{\bibfnamefont{R.~A.} \bibnamefont{Barankov}},
  \bibinfo{author}{\bibfnamefont{L.~S.} \bibnamefont{Levitov}},
  \bibnamefont{and} \bibinfo{author}{\bibfnamefont{B.~Z.}
  \bibnamefont{Spivak}}, \bibinfo{journal}{Phys. Rev. Lett.}
  \textbf{\bibinfo{volume}{93}}, \bibinfo{pages}{160401}
  (\bibinfo{year}{2004}).

\bibitem[{\citenamefont{Calabrese and Cardy}(2006)}]{cardy_quench1}
\bibinfo{author}{\bibfnamefont{P.}~\bibnamefont{Calabrese}} \bibnamefont{and}
  \bibinfo{author}{\bibfnamefont{J.}~\bibnamefont{Cardy}},
  \bibinfo{journal}{Phys. Rev. Lett.} \textbf{\bibinfo{volume}{96}},
  \bibinfo{pages}{136801} (\bibinfo{year}{2006}).

\bibitem[{\citenamefont{Calabrese and Cardy}(2007)}]{cardy_quench2}
\bibinfo{author}{\bibfnamefont{P.}~\bibnamefont{Calabrese}} \bibnamefont{and}
  \bibinfo{author}{\bibfnamefont{J.}~\bibnamefont{Cardy}}, \bibinfo{journal}{J.
  Stat. Mech: Th. and Exp.} \textbf{\bibinfo{volume}{P06008}}
  (\bibinfo{year}{2007}).

\bibitem[{\citenamefont{Sengupta et~al.}(2004)\citenamefont{Sengupta, Powell,
  and Sachdev}}]{powell}
\bibinfo{author}{\bibfnamefont{K.}~\bibnamefont{Sengupta}},
  \bibinfo{author}{\bibfnamefont{S.}~\bibnamefont{Powell}}, \bibnamefont{and}
  \bibinfo{author}{\bibfnamefont{S.}~\bibnamefont{Sachdev}},
  \bibinfo{journal}{Phys. Rev. A} \textbf{\bibinfo{volume}{69}},
  \bibinfo{pages}{053616} (\bibinfo{year}{2004}).

\bibitem[{\citenamefont{Kollath et~al.}(2007)\citenamefont{Kollath,
  L\"{a}uchli, and Altman}}]{kollath}
\bibinfo{author}{\bibfnamefont{C.}~\bibnamefont{Kollath}},
  \bibinfo{author}{\bibfnamefont{A.~M.} \bibnamefont{L\"{a}uchli}},
  \bibnamefont{and} \bibinfo{author}{\bibfnamefont{E.}~\bibnamefont{Altman}},
  \bibinfo{journal}{Phys. Rev. Lett.} \textbf{\bibinfo{volume}{98}},
  \bibinfo{pages}{180601} (\bibinfo{year}{2007}).

\bibitem[{\citenamefont{Yuzbashyan et~al.}(2005)\citenamefont{Yuzbashyan,
  Altshuler, Kuznetsov, and Enolskii}}]{yuzbashyan}
\bibinfo{author}{\bibfnamefont{E.~A.} \bibnamefont{Yuzbashyan}},
  \bibinfo{author}{\bibfnamefont{B.~L.} \bibnamefont{Altshuler}},
  \bibinfo{author}{\bibfnamefont{V.~B.} \bibnamefont{Kuznetsov}},
  \bibnamefont{and} \bibinfo{author}{\bibfnamefont{V.~Z.}
  \bibnamefont{Enolskii}}, \bibinfo{journal}{Phys. Rev. B}
  \textbf{\bibinfo{volume}{72}}, \bibinfo{pages}{220503(R)}
  (\bibinfo{year}{2005}).

\bibitem[{\citenamefont{Roux}(2009)}]{roux}
\bibinfo{author}{\bibfnamefont{G.}~\bibnamefont{Roux}}, \bibinfo{journal}{Phys.
  Rev. A} \textbf{\bibinfo{volume}{79}}, \bibinfo{pages}{021608}
  (\bibinfo{year}{2009}).

\bibitem[{\citenamefont{Gritsev
  et~al.}(2007{\natexlab{a}})\citenamefont{Gritsev, Demler, Lukin, and
  Polkovnikov}}]{gritsev_quench}
\bibinfo{author}{\bibfnamefont{V.}~\bibnamefont{Gritsev}},
  \bibinfo{author}{\bibfnamefont{E.}~\bibnamefont{Demler}},
  \bibinfo{author}{\bibfnamefont{E.}~\bibnamefont{Lukin}}, \bibnamefont{and}
  \bibinfo{author}{\bibfnamefont{A.}~\bibnamefont{Polkovnikov}},
  \bibinfo{journal}{Phys. Rev. Lett.} \textbf{\bibinfo{volume}{99}},
  \bibinfo{pages}{200404} (\bibinfo{year}{2007}{\natexlab{a}}).

\bibitem[{\citenamefont{Manmana et~al.}(2007)\citenamefont{Manmana, Wessel,
  Noack, and Muramatsu}}]{manmana}
\bibinfo{author}{\bibfnamefont{S.~R.} \bibnamefont{Manmana}},
  \bibinfo{author}{\bibfnamefont{S.}~\bibnamefont{Wessel}},
  \bibinfo{author}{\bibfnamefont{R.~M.} \bibnamefont{Noack}}, \bibnamefont{and}
  \bibinfo{author}{\bibfnamefont{A.}~\bibnamefont{Muramatsu}},
  \bibinfo{journal}{Phys. Rev. Lett.} \textbf{\bibinfo{volume}{98}},
  \bibinfo{pages}{210405} (\bibinfo{year}{2007}).

\bibitem[{\citenamefont{Mathey and Polkovnikov}(unpublished)}]{ludwig_quench}
\bibinfo{author}{\bibfnamefont{L.}~\bibnamefont{Mathey}} \bibnamefont{and}
  \bibinfo{author}{\bibfnamefont{A.}~\bibnamefont{Polkovnikov}},
  \bibinfo{journal}{arXiv:0904.2881}  (\bibinfo{year}{unpublished}).

\bibitem[{\citenamefont{Iucci and Cazalilla}(unpublished)}]{cazalilla_quench}
\bibinfo{author}{\bibfnamefont{A.}~\bibnamefont{Iucci}} \bibnamefont{and}
  \bibinfo{author}{\bibfnamefont{M.~A.} \bibnamefont{Cazalilla}},
  \bibinfo{journal}{arXiv:0903.1205}  (\bibinfo{year}{unpublished}).

\bibitem[{\citenamefont{Barmettler et~al.}(2009)\citenamefont{Barmettler, Punk,
  Gritsev, Demler, and Altman}}]{barmettler}
\bibinfo{author}{\bibfnamefont{P.}~\bibnamefont{Barmettler}},
  \bibinfo{author}{\bibfnamefont{M.}~\bibnamefont{Punk}},
  \bibinfo{author}{\bibfnamefont{V.}~\bibnamefont{Gritsev}},
  \bibinfo{author}{\bibfnamefont{E.}~\bibnamefont{Demler}}, \bibnamefont{and}
  \bibinfo{author}{\bibfnamefont{E.}~\bibnamefont{Altman}},
  \bibinfo{journal}{Phys. Rev. Lett.} \textbf{\bibinfo{volume}{102}},
  \bibinfo{pages}{130603} (\bibinfo{year}{2009}).

\bibitem[{\citenamefont{Polkovnikov}(2005)}]{ap_adiabatic}
\bibinfo{author}{\bibfnamefont{A.}~\bibnamefont{Polkovnikov}},
  \bibinfo{journal}{Phys. Rev. B} \textbf{\bibinfo{volume}{72}},
  \bibinfo{pages}{R161201} (\bibinfo{year}{2005}).

\bibitem[{\citenamefont{Zurek et~al.}(2005)\citenamefont{Zurek, Dorner, and
  Zoller}}]{zurek_adiabatic}
\bibinfo{author}{\bibfnamefont{W.~H.} \bibnamefont{Zurek}},
  \bibinfo{author}{\bibfnamefont{U.}~\bibnamefont{Dorner}}, \bibnamefont{and}
  \bibinfo{author}{\bibfnamefont{P.}~\bibnamefont{Zoller}},
  \bibinfo{journal}{Phys. Rev. Lett.} \textbf{\bibinfo{volume}{95}},
  \bibinfo{pages}{105701} (\bibinfo{year}{2005}).

\bibitem[{\citenamefont{Dziarmaga}(2005)}]{jacek}
\bibinfo{author}{\bibfnamefont{J.}~\bibnamefont{Dziarmaga}},
  \bibinfo{journal}{Phys. Rev. Lett.} \textbf{\bibinfo{volume}{95}},
  \bibinfo{pages}{245701} (\bibinfo{year}{2005}).

\bibitem[{\citenamefont{Cherng and Levitov}(2006)}]{cherng-levitov}
\bibinfo{author}{\bibfnamefont{R.~W.} \bibnamefont{Cherng}} \bibnamefont{and}
  \bibinfo{author}{\bibfnamefont{L.~S.} \bibnamefont{Levitov}},
  \bibinfo{journal}{Phys. Rev. A} \textbf{\bibinfo{volume}{73}},
  \bibinfo{pages}{043614} (\bibinfo{year}{2006}).

\bibitem[{\citenamefont{Polkovnikov and Gritsev}(2008)}]{pg_np}
\bibinfo{author}{\bibfnamefont{A.}~\bibnamefont{Polkovnikov}} \bibnamefont{and}
  \bibinfo{author}{\bibfnamefont{V.}~\bibnamefont{Gritsev}},
  \bibinfo{journal}{Nature Phys.} \textbf{\bibinfo{volume}{4}},
  \bibinfo{pages}{477} (\bibinfo{year}{2008}).

\bibitem[{\citenamefont{Altland and Gurarie}(2008)}]{ag}
\bibinfo{author}{\bibfnamefont{A.}~\bibnamefont{Altland}} \bibnamefont{and}
  \bibinfo{author}{\bibfnamefont{V.}~\bibnamefont{Gurarie}},
  \bibinfo{journal}{Phys. Rev. Lett.} \textbf{\bibinfo{volume}{100}},
  \bibinfo{pages}{063602} (\bibinfo{year}{2008}).

\bibitem[{\citenamefont{Barankov and Polkovnikov}(2008)}]{bp_optim}
\bibinfo{author}{\bibfnamefont{R.}~\bibnamefont{Barankov}} \bibnamefont{and}
  \bibinfo{author}{\bibfnamefont{A.}~\bibnamefont{Polkovnikov}},
  \bibinfo{journal}{Phys. Rev. Lett.} \textbf{\bibinfo{volume}{101}},
  \bibinfo{pages}{076801} (\bibinfo{year}{2008}).

\bibitem[{\citenamefont{Bistritzer and Altman}(2007)}]{rafi}
\bibinfo{author}{\bibfnamefont{R.}~\bibnamefont{Bistritzer}} \bibnamefont{and}
  \bibinfo{author}{\bibfnamefont{E.}~\bibnamefont{Altman}},
  \bibinfo{journal}{PNAS} \textbf{\bibinfo{volume}{104}}, \bibinfo{pages}{9955}
  (\bibinfo{year}{2007}).

\bibitem[{\citenamefont{De\:\:Grandi et~al.}(2008)\citenamefont{De\:\:Grandi,
  Barankov, and Polkovnikov}}]{claudia}
\bibinfo{author}{\bibfnamefont{C.}~\bibnamefont{De\:\:Grandi}},
  \bibinfo{author}{\bibfnamefont{R.~A.} \bibnamefont{Barankov}},
  \bibnamefont{and}
  \bibinfo{author}{\bibfnamefont{A.}~\bibnamefont{Polkovnikov}},
  \bibinfo{journal}{Phys. Rev. Lett.} \textbf{\bibinfo{volume}{101}},
  \bibinfo{pages}{230402} (\bibinfo{year}{2008}).

\bibitem[{\citenamefont{Sengupta et~al.}(2008)\citenamefont{Sengupta, Sen, and
  Mondal}}]{sengupta_sen2008a}
\bibinfo{author}{\bibfnamefont{K.}~\bibnamefont{Sengupta}},
  \bibinfo{author}{\bibfnamefont{D.}~\bibnamefont{Sen}}, \bibnamefont{and}
  \bibinfo{author}{\bibfnamefont{S.}~\bibnamefont{Mondal}},
  \bibinfo{journal}{Phys. Rev. Lett.} \textbf{\bibinfo{volume}{100}},
  \bibinfo{pages}{077204} (\bibinfo{year}{2008}).

\bibitem[{\citenamefont{Sen et~al.}(2008)\citenamefont{Sen, Sengupta, and
  Mondal}}]{sengupta_sen2008b}
\bibinfo{author}{\bibfnamefont{D.}~\bibnamefont{Sen}},
  \bibinfo{author}{\bibfnamefont{K.}~\bibnamefont{Sengupta}}, \bibnamefont{and}
  \bibinfo{author}{\bibfnamefont{S.}~\bibnamefont{Mondal}},
  \bibinfo{journal}{Phys. Rev. Lett.} \textbf{\bibinfo{volume}{101}},
  \bibinfo{pages}{016806} (\bibinfo{year}{2008}).

\bibitem[{\citenamefont{Altland et~al.}(2009)\citenamefont{Altland, Gurarie,
  Kriecherbauer, and Polkovnikov}}]{agkp}
\bibinfo{author}{\bibfnamefont{A.}~\bibnamefont{Altland}},
  \bibinfo{author}{\bibfnamefont{V.}~\bibnamefont{Gurarie}},
  \bibinfo{author}{\bibfnamefont{T.}~\bibnamefont{Kriecherbauer}},
  \bibnamefont{and}
  \bibinfo{author}{\bibfnamefont{A.}~\bibnamefont{Polkovnikov}},
  \bibinfo{journal}{Phys. Rev. A} \textbf{\bibinfo{volume}{79}},
  \bibinfo{pages}{042703} (\bibinfo{year}{2009}).

\bibitem[{\citenamefont{Divakaran et~al.}(2009)\citenamefont{Divakaran,
  Mukherjee, Dutta, and Sen}}]{dutta2009}
\bibinfo{author}{\bibfnamefont{U.}~\bibnamefont{Divakaran}},
  \bibinfo{author}{\bibfnamefont{V.}~\bibnamefont{Mukherjee}},
  \bibinfo{author}{\bibfnamefont{A.}~\bibnamefont{Dutta}}, \bibnamefont{and}
  \bibinfo{author}{\bibfnamefont{D.}~\bibnamefont{Sen}}, \bibinfo{journal}{J.
  Stat. Mech.} p. \bibinfo{pages}{P02007} (\bibinfo{year}{2009}).

\bibitem[{\citenamefont{Chowdhury et~al.}(unpublished)\citenamefont{Chowdhury,
  Divakaran, and Dutta}}]{dutta2009a}
\bibinfo{author}{\bibfnamefont{D.}~\bibnamefont{Chowdhury}},
  \bibinfo{author}{\bibfnamefont{U.}~\bibnamefont{Divakaran}},
  \bibnamefont{and} \bibinfo{author}{\bibfnamefont{A.}~\bibnamefont{Dutta}},
  \bibinfo{journal}{arXiv:0906.1161}  (\bibinfo{year}{unpublished}).

\bibitem[{\citenamefont{Sengupta and Sen}(unpublished)}]{sengupta_sen2009}
\bibinfo{author}{\bibfnamefont{K.}~\bibnamefont{Sengupta}} \bibnamefont{and}
  \bibinfo{author}{\bibfnamefont{D.}~\bibnamefont{Sen}},
  \bibinfo{journal}{arXiv:0904.1059}  (\bibinfo{year}{unpublished}).

\bibitem[{\citenamefont{Itin and T\"orm\"a}(unpublished)}]{itin}
\bibinfo{author}{\bibfnamefont{A.}~\bibnamefont{Itin}} \bibnamefont{and}
  \bibinfo{author}{\bibfnamefont{P.}~\bibnamefont{T\"orm\"a}},
  \bibinfo{journal}{preprint: arXiv:0901.4778}  (\bibinfo{year}{unpublished}).

\bibitem[{\citenamefont{Rossini et~al.}(2009)\citenamefont{Rossini, Silva,
  Mussardo, and Santoro}}]{silva2009}
\bibinfo{author}{\bibfnamefont{D.}~\bibnamefont{Rossini}},
  \bibinfo{author}{\bibfnamefont{A.}~\bibnamefont{Silva}},
  \bibinfo{author}{\bibfnamefont{G.}~\bibnamefont{Mussardo}}, \bibnamefont{and}
  \bibinfo{author}{\bibfnamefont{G.~E.} \bibnamefont{Santoro}},
  \bibinfo{journal}{Phys. Rev. Lett.} \textbf{\bibinfo{volume}{102}},
  \bibinfo{pages}{127204} (\bibinfo{year}{2009}).

\bibitem[{\citenamefont{Pollmann et~al.}(unpublished)\citenamefont{Pollmann,
  Mukerjee, Green, and Moore}}]{moore2009}
\bibinfo{author}{\bibfnamefont{F.}~\bibnamefont{Pollmann}},
  \bibinfo{author}{\bibfnamefont{S.}~\bibnamefont{Mukerjee}},
  \bibinfo{author}{\bibfnamefont{A.~G.} \bibnamefont{Green}}, \bibnamefont{and}
  \bibinfo{author}{\bibfnamefont{J.~E.} \bibnamefont{Moore}},
  \bibinfo{journal}{arXiv:0907.3206}  (\bibinfo{year}{unpublished}).

\bibitem[{cla()}]{claudia_quench}
\bibinfo{note}{C. De\:\:Grandi, V. Gritsev, A. Polkovnikov, arXiv:0909.5181,
  arXiv:0910.0876 (unpublished)}.

\bibitem[{\citenamefont{Landau}(1932)}]{lz1}
\bibinfo{author}{\bibfnamefont{L.}~\bibnamefont{Landau}},
  \bibinfo{journal}{Phys. Z. Sowj.} \textbf{\bibinfo{volume}{2}},
  \bibinfo{pages}{46} (\bibinfo{year}{1932}).

\bibitem[{\citenamefont{Zener}(1932)}]{lz2}
\bibinfo{author}{\bibfnamefont{C.}~\bibnamefont{Zener}},
  \bibinfo{journal}{Proc. R. Soc.} \textbf{\bibinfo{volume}{137}},
  \bibinfo{pages}{696} (\bibinfo{year}{1932}).

\bibitem[{\citenamefont{Kibble}(1976)}]{kz1}
\bibinfo{author}{\bibfnamefont{T.~W.~B.} \bibnamefont{Kibble}},
  \bibinfo{journal}{J. Phys. A} \textbf{\bibinfo{volume}{9}},
  \bibinfo{pages}{1387} (\bibinfo{year}{1976}).

\bibitem[{\citenamefont{Zurek}(1996)}]{kz2}
\bibinfo{author}{\bibfnamefont{W.~H.} \bibnamefont{Zurek}},
  \bibinfo{journal}{Phys. Rep.} \textbf{\bibinfo{volume}{276}},
  \bibinfo{pages}{177} (\bibinfo{year}{1996}).

\bibitem[{\citenamefont{Rigolin et~al.}(2008)\citenamefont{Rigolin, Ortiz, and
  Ponce}}]{ortiz2008}
\bibinfo{author}{\bibfnamefont{G.}~\bibnamefont{Rigolin}},
  \bibinfo{author}{\bibfnamefont{G.}~\bibnamefont{Ortiz}}, \bibnamefont{and}
  \bibinfo{author}{\bibfnamefont{V.~H.} \bibnamefont{Ponce}},
  \bibinfo{journal}{Phys. Rev. A} \textbf{\bibinfo{volume}{78}},
  \bibinfo{pages}{052508} (\bibinfo{year}{2008}).

\bibitem[{\citenamefont{Landau and Lifshitz}(1981)}]{LL3}
\bibinfo{author}{\bibfnamefont{L.}~\bibnamefont{Landau}} \bibnamefont{and}
  \bibinfo{author}{\bibfnamefont{E.}~\bibnamefont{Lifshitz}},
  \emph{\bibinfo{title}{Quantum Mechanics: Non-Relativistic Theory, Vol. 3}}
  (\bibinfo{publisher}{Butterworth-Heinemann}, \bibinfo{year}{1981}).

\bibitem[{\citenamefont{Shankar}(1994)}]{shankar}
\bibinfo{author}{\bibfnamefont{R.}~\bibnamefont{Shankar}},
  \emph{\bibinfo{title}{Principles of Quantum Mechanics}}
  (\bibinfo{publisher}{Springer, New York}, \bibinfo{year}{1994}).

\bibitem[{\citenamefont{Vitanov and Garraway}(1996)}]{vitanov_96}
\bibinfo{author}{\bibfnamefont{N.~V.} \bibnamefont{Vitanov}} \bibnamefont{and}
  \bibinfo{author}{\bibfnamefont{B.~M.} \bibnamefont{Garraway}},
  \bibinfo{journal}{Phys. Rev. A} \textbf{\bibinfo{volume}{53}},
  \bibinfo{pages}{4288} (\bibinfo{year}{1996}).

\bibitem[{\citenamefont{Vitanov}(1999)}]{vitanov_99}
\bibinfo{author}{\bibfnamefont{N.~V.} \bibnamefont{Vitanov}},
  \bibinfo{journal}{Phys. Rev. A} \textbf{\bibinfo{volume}{59}},
  \bibinfo{pages}{988} (\bibinfo{year}{1999}).

\bibitem[{\citenamefont{Gelmont et~al.}(1976)\citenamefont{Gelmont,
  Ivanovomskii, and Tsidilkovskii}}]{gelmont}
\bibinfo{author}{\bibfnamefont{B.~L.} \bibnamefont{Gelmont}},
  \bibinfo{author}{\bibfnamefont{V.~I.} \bibnamefont{Ivanovomskii}},
  \bibnamefont{and} \bibinfo{author}{\bibfnamefont{I.~M.}
  \bibnamefont{Tsidilkovskii}}, \bibinfo{journal}{Uspekhi Fizicheskikh Nauk}
  \textbf{\bibinfo{volume}{120}}, \bibinfo{pages}{337} (\bibinfo{year}{1976}).

\bibitem[{\citenamefont{Neto et~al.}(2009)\citenamefont{Neto, Guinea, Peres,
  and andA. K.~Geim}}]{antonio_rmp}
\bibinfo{author}{\bibfnamefont{A.~H.~C.} \bibnamefont{Neto}},
  \bibinfo{author}{\bibfnamefont{F.}~\bibnamefont{Guinea}},
  \bibinfo{author}{\bibfnamefont{N.~M.~R.} \bibnamefont{Peres}},
  \bibnamefont{and} \bibinfo{author}{\bibfnamefont{K.~S.~N.} \bibnamefont{andA.
  K.~Geim}}, \bibinfo{journal}{Rev. Mod. Phys.} \textbf{\bibinfo{volume}{81}},
  \bibinfo{pages}{109} (\bibinfo{year}{2009}).

\bibitem[{\citenamefont{Giamarchi}(2004)}]{giamarchi_book}
\bibinfo{author}{\bibfnamefont{T.}~\bibnamefont{Giamarchi}},
  \emph{\bibinfo{title}{Quantum Physics in One Dimension}}
  (\bibinfo{publisher}{Clarendon Press}, \bibinfo{address}{Oxford},
  \bibinfo{year}{2004}).

\bibitem[{\citenamefont{Sachdev}(1999)}]{subir}
\bibinfo{author}{\bibfnamefont{S.}~\bibnamefont{Sachdev}},
  \emph{\bibinfo{title}{Quantum Phase Transitions}}
  (\bibinfo{publisher}{Cambridge University Press},
  \bibinfo{address}{Cambridge}, \bibinfo{year}{1999}).

\bibitem[{gir()}]{girardeaus}
\bibinfo{note}{M.~Girardeau, J. Math .Phys. {\bf 1}, 516 (1960); C.~N.~Yang and
  Y.~P.~Yang, J. Math .Phys. {\bf 10}, 1115 (1969); L.~Tonks, Phys. Rev. {\bf
  50}, 955 (1936).}

\bibitem[{\citenamefont{Paredes et~al.}(2004)\citenamefont{Paredes, Widera,
  Murg, Mandel, F\"olling, Cirac, Shlyapnikov, Hansch, and
  Bloch}}]{bloch_tonks}
\bibinfo{author}{\bibfnamefont{B.}~\bibnamefont{Paredes}},
  \bibinfo{author}{\bibfnamefont{A.}~\bibnamefont{Widera}},
  \bibinfo{author}{\bibfnamefont{V.}~\bibnamefont{Murg}},
  \bibinfo{author}{\bibfnamefont{O.}~\bibnamefont{Mandel}},
  \bibinfo{author}{\bibfnamefont{S.}~\bibnamefont{F\"olling}},
  \bibinfo{author}{\bibfnamefont{I.}~\bibnamefont{Cirac}},
  \bibinfo{author}{\bibfnamefont{G.~V.} \bibnamefont{Shlyapnikov}},
  \bibinfo{author}{\bibfnamefont{T.~W.} \bibnamefont{Hansch}},
  \bibnamefont{and} \bibinfo{author}{\bibfnamefont{I.}~\bibnamefont{Bloch}},
  \bibinfo{journal}{Nature} \textbf{\bibinfo{volume}{429}},
  \bibinfo{pages}{277} (\bibinfo{year}{2004}).

\bibitem[{\citenamefont{Polkovnikov}(2008)}]{ap_heat}
\bibinfo{author}{\bibfnamefont{A.}~\bibnamefont{Polkovnikov}},
  \bibinfo{journal}{Phys. Rev. Lett.} \textbf{\bibinfo{volume}{101}},
  \bibinfo{pages}{220402} (\bibinfo{year}{2008}).

\bibitem[{\citenamefont{Barankov and Polkovnikov}(unpublished)}]{diag_ent}
\bibinfo{author}{\bibfnamefont{R.}~\bibnamefont{Barankov}} \bibnamefont{and}
  \bibinfo{author}{\bibfnamefont{A.}~\bibnamefont{Polkovnikov}},
  \bibinfo{journal}{arXiv:0806.2862}  (\bibinfo{year}{unpublished}).

\bibitem[{\citenamefont{Gritsev
  et~al.}(2007{\natexlab{b}})\citenamefont{Gritsev, Polkovnikov, and
  Demler}}]{vova_lr}
\bibinfo{author}{\bibfnamefont{V.}~\bibnamefont{Gritsev}},
  \bibinfo{author}{\bibfnamefont{A.}~\bibnamefont{Polkovnikov}},
  \bibnamefont{and} \bibinfo{author}{\bibfnamefont{E.}~\bibnamefont{Demler}},
  \bibinfo{journal}{Phys. Rev. B} \textbf{\bibinfo{volume}{75}},
  \bibinfo{pages}{174511} (\bibinfo{year}{2007}{\natexlab{b}}).

\bibitem[{\citenamefont{B\"uchler et~al.}(2003)\citenamefont{B\"uchler,
  Blatter, and Zwerger}}]{buchler}
\bibinfo{author}{\bibfnamefont{H.}~\bibnamefont{B\"uchler}},
  \bibinfo{author}{\bibfnamefont{G.}~\bibnamefont{Blatter}}, \bibnamefont{and}
  \bibinfo{author}{\bibfnamefont{W.}~\bibnamefont{Zwerger}},
  \bibinfo{journal}{Phys. Rev. Lett.} \textbf{\bibinfo{volume}{90}},
  \bibinfo{pages}{130401} (\bibinfo{year}{2003}).

\bibitem[{\citenamefont{Damski and Zurek}(2006)}]{damski_zurek}
\bibinfo{author}{\bibfnamefont{B.}~\bibnamefont{Damski}} \bibnamefont{and}
  \bibinfo{author}{\bibfnamefont{W.~H.} \bibnamefont{Zurek}},
  \bibinfo{journal}{Phys. Rev. A} \textbf{\bibinfo{volume}{73}},
  \bibinfo{pages}{063405} (\bibinfo{year}{2006}).

\bibitem[{\citenamefont{Altman and Vishwanath}(2005)}]{ehud_ashvin}
\bibinfo{author}{\bibfnamefont{E.}~\bibnamefont{Altman}} \bibnamefont{and}
  \bibinfo{author}{\bibfnamefont{A.}~\bibnamefont{Vishwanath}},
  \bibinfo{journal}{Phys. Rev. Lett.} \textbf{\bibinfo{volume}{95}},
  \bibinfo{pages}{110404} (\bibinfo{year}{2005}).

\bibitem[{\citenamefont{Gu and Lin}(2009)}]{gu}
\bibinfo{author}{\bibfnamefont{S.-J.} \bibnamefont{Gu}} \bibnamefont{and}
  \bibinfo{author}{\bibfnamefont{H.-Q.} \bibnamefont{Lin}},
  \bibinfo{journal}{Europhys. Lett.} \textbf{\bibinfo{volume}{87}},
  \bibinfo{pages}{10003} (\bibinfo{year}{2009}).

\bibitem[{\citenamefont{Polkovnikov}(unpublished)}]{ap_wig}
\bibinfo{author}{\bibfnamefont{A.}~\bibnamefont{Polkovnikov}},
  \bibinfo{journal}{arXiv:0905.3384}  (\bibinfo{year}{unpublished}).

\bibitem[{\citenamefont{Sotiriadis
  et~al.}(unpublished)\citenamefont{Sotiriadis, Calabrese, and
  Cardy}}]{cardy_temp}
\bibinfo{author}{\bibfnamefont{S.}~\bibnamefont{Sotiriadis}},
  \bibinfo{author}{\bibfnamefont{P.}~\bibnamefont{Calabrese}},
  \bibnamefont{and} \bibinfo{author}{\bibfnamefont{J.}~\bibnamefont{Cardy}},
  \bibinfo{journal}{arXiv:0903.0895}  (\bibinfo{year}{unpublished}).

\bibitem[{\citenamefont{Abrikosov et~al.}(1975)\citenamefont{Abrikosov,
  Gor'kov, and Dzyaloshinski}}]{abrikosov}
\bibinfo{author}{\bibfnamefont{A.~A.} \bibnamefont{Abrikosov}},
  \bibinfo{author}{\bibfnamefont{L.~P.} \bibnamefont{Gor'kov}},
  \bibnamefont{and} \bibinfo{author}{\bibfnamefont{I.~E.}
  \bibnamefont{Dzyaloshinski}}, \emph{\bibinfo{title}{Methods of quantum field
  theory in statistical physics}} (\bibinfo{publisher}{edited by R. A.
  Silverman, Courier Dover Publications}, \bibinfo{year}{1975}).

\end{thebibliography}

\end{document}